\documentclass[12pt]{article}
\usepackage{graphicx,psfrag,multirow}

\newcommand \bra[1]{\left< {#1} \,\right\vert}
\newcommand \ket[1]{\left\vert\, {#1} \, \right>}
\newcommand \braket[2]{\hbox{$\left< {#1} \,\vrule\, {#2} \right>$}}
\newcommand{\bea}{\begin{eqnarray}}
\newcommand{\eea}{\end{eqnarray}}
\newcommand{\simgt}{\hbox{ \raise3pt\hbox to 0pt{$>$}\raise-3pt\hbox{$\sim$} }}
\newcommand{\simlt}{\hbox{ \raise3pt\hbox to 0pt{$<$}\raise-3pt\hbox{$\sim$} }}

\newcommand{\clfn}{\setcounter{footnote}{0}}

\begin{document}
\begin{titlepage}
\title{Improved Perturbative QCD Approach to\\
the Bottomonium Spectrum \vspace{2cm}}
\author{S.~Recksiegel$^1$ and Y.~Sumino$^2$
\\
\\ $^1$ Theory Group, KEK\\
Tsukuba, Ibaraki, 305-0801 Japan\\
\\ $^2$ Department of Physics, Tohoku University\\
Sendai, 980-8578 Japan
}
\date{}
\maketitle
\thispagestyle{empty}
\vspace{-5truein}
\begin{flushright}
{\bf hep-ph/0207005}\\
{\bf TU--659}\\
{\bf KEK--TH--827}\\
{\bf June 2002}
\end{flushright}
\vspace{4.5truein}
\begin{abstract}
\noindent
{\small
Recently it has been shown that the gross structure of the
bottomonium spectrum is reproduced reasonably well within the
non--relativistic boundstate theory based on perturbative QCD.
In that calculation, however, the fine splittings 
and the $S$--$P$ level splittings 
are predicted to be considerably narrower than the 
corresponding experimental values.
We investigate the bottomonium spectrum within a specific
framework based on perturbative QCD, which incorporates
all the corrections up to ${\cal O}(\alpha_S^5 m_b)$ and 
${\cal O}(\alpha_S^4 m_b)$, respectively, 
in the computations of the fine splittings and the
$S$--$P$ splittings.
We find that the agreement with the experimental data for the
fine splittings improves drastically
due to an enhancement of the wave functions close to 
the origin as compared to the Coulomb wave functions.
The agreement of the $S$--$P$ splittings with the experimental
data also becomes better.
We find that natural scales of 
the fine splittings and the $S$--$P$ splittings
are larger than those of the boundstates themselves.
On the other hand,
the predictions of the level spacings between consecutive
principal quantum numbers 
depend rather strongly on the scale $\mu$ of the
operator $\propto C_A/(m_b r^2)$.
The agreement of the whole spectrum 
with the experimental data is much better than the previous
predictions when $\mu \simeq 3$--4~GeV for $\alpha_S(M_Z)=0.1181$.
There seems to be a phenomenological preference for some suppression
mechanism for the above operator.
}
\end{abstract}
\vfil

\end{titlepage}
  
\section{Introduction}

For a long time most successful theoretical
approaches to study the heavy quarkonium spectra have been
those based on various phenomenological potential models.
These model approaches have been capable
not only of reproducing the charmonium and
bottomonium spectra to a high accuracy, but also of explaining
various other properties of heavy quarkonia such as their
transition rates and decay rates.
Through this success,
these phenomenological models have established, essentially, 
that the heavy quarkonium states can be described well as
non--relativistic boundstate systems;
see e.g.\ \cite{eq} for one of the most recent analyses.
On the other hand, the problem of the phenomenological approaches is that
it is difficult to improve the theoretical predictions systematically,
and that it is difficult to relate the parameters of the models
to the fundamental theory.

Recently there has been new progress in explaining these
heavy quarkonium spectra within the framework of
non-relativistic boundstate theory based on perturbative QCD.
It has been shown that, by incorporating the cancellation
of ${\cal O}(\Lambda_{\rm QCD})$ renormalons contained in the pole mass
and the static QCD potential,
the gross structure of the bottomonium spectrum is
reproduced reasonably well by the perturbative computation of
the spectrum up to ${\cal O}(1/c^2)={\cal O}(\alpha_S^4 m_b)$
\cite{bsv1,bsv2}.
Furthermore, it has been shown that the static QCD potential, 
calculated in a series expansion in $\alpha_S$ up to ${\cal O}(\alpha_S^3)$,
agrees well with typical phenomenological potentials
in the region relevant to bottomonium spectroscopy,
once the ${\cal O}(\Lambda_{\rm QCD})$ renormalon contained in
the QCD potential is cancelled against that contained in the
$b$--quark pole mass \cite{paper1,RS1}.
Since the static QCD potential calculated by
lattice simulations is consistent with the phenomenological 
potentials in this region \cite{bsw,ukqcd}, 
all these calculations are consistent with one another.
See also \cite{necco-sommer}, which made a direct comparison of lattice
results and perturbative predictions of the QCD potential
in a renormalon--subtracted scheme 
and found the same consistency.

The key concept
which led to these new results from perturbative QCD, is the following.
Conventionally, theoretical calculations of the energy levels
of a non--relativistic quark--antiquark boundstate
closely followed that of a QED boundstate such as positronium:
it starts from the natural picture that, when an electron and a 
positron are at rest and 
far apart from each other, they tend to be free particles and the
total energy of the system is given by the sum of the energies of
the two particles (pole masses);
as the electron and positron approach each other, the
energy of the system decreases due to the 
negative potential energy, so that
the total energy of the boundstate is given as 
the sum of the pole masses minus the binding energy.
When the calculation along the same line
was applied to the quark--antiquark
system, however, the perturbative expansion of the boundstate energy
turned out to be poorly convergent, due to the contributions from
infrared (IR) gluons with wave--lengths of order $\Lambda_{\rm QCD}^{-1}$.
We can regard this as reflecting the invalidity of the free quark picture
when the quark and antiquark are far apart from each other.
On the other hand, intuitively 
we expect that there should be a way to calculate
the boundstate energy in which the contributions of IR gluons can
be mostly eliminated.
This is because when the boundstate size is sufficiently smaller
than $\Lambda_{\rm QCD}^{-1}$, IR gluons cannot resolve the colour
charges of the constituent particles, so that they decouple from
this colour--singlet system.
Indeed this idea was theoretically validated in the language of renormalons
and their cancellation \cite{renormalon1,renormalon2}.
As a result, the convergence of the perturbative expansion improved
dramatically, extending the predictive power of perturbative QCD
beyond what could be achieved before.

Nevertheless,
there remain some problems regarding the above prediction of the
bottomonium spectrum from perturbative QCD.
Among them, especially interesting is 
the problem that the fine structure in the $1P_j$ levels
as well as the splittings between the $2S$ and $1P_j$ states
are predicted to be considerably narrower than the 
corresponding experimental values.
One may think that the level of
agreement of the theoretical prediction
with the experimental data is still consistent within errors:
according to an estimate based on next--to--leading order
renormalons, each energy level has a theoretical uncertainty of order 
$\Lambda_{\rm QCD}^3\cdot a^2$ ($a$ is the typical size of the corresponding
boundstate) which may be
comparable in size to the $2S$--$1P_j$ splittings
and may be much larger than the fine splittings in the $1P_j$ levels;
compare the error estimate in \cite{bsv2}.
One should note, however, that if we calculate these level splittings 
instead of the individual energy levels, the 
${\cal O}(\Lambda_{\rm QCD}^3)$ renormalons should get largely cancelled
when we take the differences of the energy levels.
Hence, the theoretical uncertainties of the splittings can be much
smaller than those of the individual energy levels, 
and the disagreement between the theoretical prediction
and the experimental data may be quite serious.
It is important to clarify whether 
it is possible to resolve these disagreements in the fine splittings
and the $S$--$P$ splittings within the context of
perturbative QCD, e.g.\ by including higher--order corrections,
or whether we need to take into account specific non--perturbative 
contributions for this purpose.

In this paper we investigate this problem of the
fine splittings and the $S$--$P$ splittings in the bottomonium spectrum
within a specific framework based on perturbative QCD.
We note that these splittings have been successfully reproduced
by the phenomenological potential models, and
that a connection between the static QCD potential and
phenomenological potentials has been elucidated in \cite{paper1,RS1}.
In order to take advantage of these results,
we develop a framework which enables detailed comparison of the
predictions of the
phenomenological models and of perturbative QCD. 
We also incorporate some of the
higher--order corrections to the non--relativistic Hamiltonian
of the quark--antiquark system which have not been
included in the analysis \cite{bsv2}.

The basic theoretical ingredients of our analysis are as follows:
(1) We take into account the cancellation of the ${\cal O}(\Lambda_{\rm QCD})$
leading renormalons  by reexpressing the $b$--quark
pole mass in terms of the $\overline{\rm MS}$ mass.
(2) We take a specific scheme for the perturbative expansion such 
that all the corrections up to
${\cal O}(\alpha_S^4 m_b)$ and ${\cal O}(\alpha_S^5 m_b)$ 
are incorporated in the calculation of the $S$--$P$ splittings and the
fine splittings, respectively.
Furthermore, some of the
higher--order corrections, which appear to be important for these
observables, are incorporated.

The organization of the paper is as follows.
In Sec.~2 we present the framework of our calculation.
We examine the energy levels and the wave functions of our zeroth--order 
Hamiltonian in Sec.~3.
The analysis of the fine splittings is given in Sec.~4
and that of the $S$--$P$ level splittings in Sec.~5.
Then, in Sec.~6, we compare the whole structure of the bottomonium spectrum 
given by our prediction, by other theoretical predictions and by the
experimental data.
Concluding remarks are given in Sec.~7.
We derive a formula useful for our analysis in the Appendix.


\section{Framework of Calculation}
\label{s2}

\subsection{Hamiltonian up to \boldmath{${\cal O}(1/c^2)$}}

We first recall 
the non--relativistic Hamiltonian of a 
quark and antiquark pair given in the series expansion in
$1/c$ up to ${\cal O}(1/c^2)$ which is determined from 
perturbative QCD.
(See e.g.\ \cite{yn3,ps,py,htmy}.)
Considering an application to the bottomonium states,
we assume the quark (antiquark) to be the
$b$--quark ($\bar{b}$-quark).
The Hamiltonian is given by
\bea
&&
H = H_0 + U + W_A + W_{NA} .
\label{hamiltonian}
\eea
We choose the zeroth-order 
part of the Hamiltonian to be 
\bea
&&
H_0 = 2 m_b + \frac{\vec{p}\,^2}{m_b} + V_{\rm QCD}(r) ,
\eea
where $m_b$ is the pole mass of the $b$--quark, and
$V_{\rm QCD}(r)$ denotes the static QCD potential up to
${\cal O}(\alpha_S^3)$.
This choice differs from the usual zeroth--order Hamiltonian
of the $1/c$--expansion, since $H_0$ also includes the 
${\cal O}(\alpha_S^2)={\cal O}(1/c)$ and 
${\cal O}(\alpha_S^3)={\cal O}(1/c^2)$ terms of
the QCD potential.
Other operators of Eq.~(\ref{hamiltonian})
are treated as perturbations to $H_0$, all of which are
${\cal O}(1/c^2)$ in the usual order counting in $1/c$-expansion.
$U + W_A$
constitutes the ${\cal O}(1/c^2)$ part of the Breit Hamiltonian
known from QED boundstate theory, where
the spin--dependent operator is given by
\bea
&&
U = U_{LS}\, \, \vec{L}\cdot\vec{S}  
+ U_S \, \,
\biggl[ {S^2} - 3 \frac{(\vec{S}\cdot\vec{r})^2}{r^2} \biggr] 
+ U_0 \,\, (2S^2-3) \delta^3(\vec{r}) ,
\\&&
U_{LS} = \frac{3 C_F \alpha_S^{(n_l)}}{2m_b^2r^3} ,
\qquad
U_S = - \frac{C_F \alpha_S^{(n_l)}}{2m_b^2r^3} ,
\qquad
U_0 = \frac{2\pi C_F \alpha_S^{(n_l)}}{3m_b^2} ,
\eea
and the spin--independent operator is given by
\bea
&&
W_A =  - \frac{\vec{p}\,^4}{4m_b^3} \,
+ \frac{\pi C_F \alpha_S^{(n_l)}}{m_b^2} \, \delta^3(\vec{r})
- \frac{C_F \alpha_S^{(n_l)}}{2m_b^2r} \biggl(
\vec{p}\,^2 + \frac{1}{r^2} r_i r_j p_j p_i \biggr) .
\label{WA}
\eea
On the other hand,
\bea
&&
W_{NA} =
- \frac{C_A C_F (\alpha_S^{(n_l)})^2}{2m_br^2}
\eea
represents the operator characteristic to the non--abelian gauge theory.
In this paper, unless the argument is specified explicitly,
$\alpha_S^{(n_l)}$ denotes the strong coupling
constant renormalized at the renormalization scale $\mu$, defined
in the $\overline{\rm MS}$ scheme with $n_l$ active flavours,
i.e.\  $\alpha_S^{(n_l)}\equiv \alpha_S^{(n_l)}(\mu)$;
$C_F = 4/3$ and $C_A =3$ are the colour factors;
$\vec{L}$ and $\vec{S}$ are, respectively, the orbital--angular momentum
and the total spin of the quark--antiquark pair.
For the $b\bar{b}$ system, $n_l=4$.

\subsection{Improved Potential}

\begin{figure}
\begin{center}
\psfrag{R}{$r\,[{\rm GeV}^{-1}]$}\psfrag{Energy}{$E_{\rm imp}(r)$ [GeV]}
\psfrag{RG}{RG--improved potential}
\psfrag{QCD}{$E^{b\bar{b}}_{\rm tot}(r)$}
\psfrag{linear}{linear}\psfrag{extrap}{extrapolation}
\psfrag{rUV}{$r_{\rm UV}=0.5\,{\rm GeV}^{-1}$}
\psfrag{rIR}{$r_{\rm IR}\,\,=4.5\,{\rm GeV}^{-1}$}
\includegraphics[width=12cm]{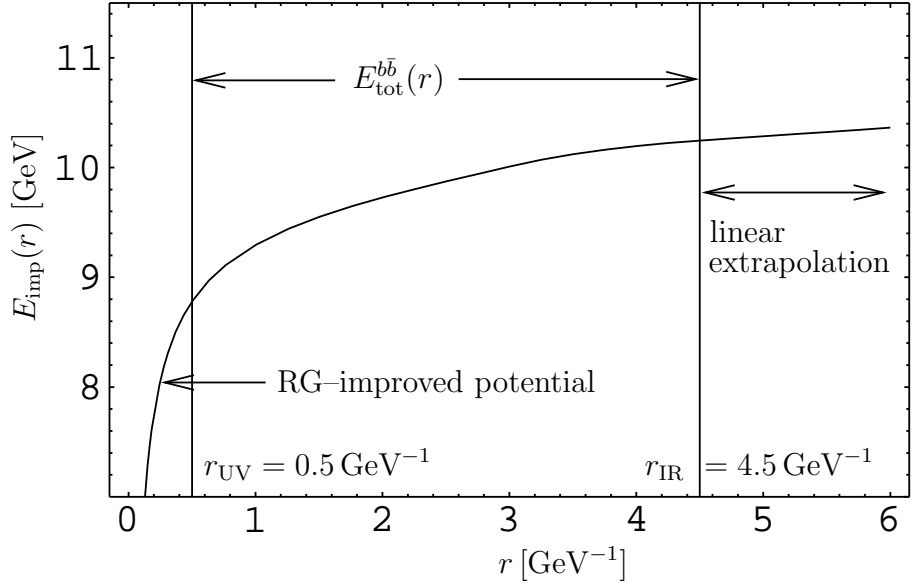}
\end{center}
\vspace*{-.5cm}
\caption{\small
Construction of the improved potential. \label{potential}}
\end{figure}

In our analysis of the bottomonium spectrum and wave functions,
we use an improved ``potential'' $E_{\rm imp}(r)$
instead of $2m_b + V_{\rm QCD}(r)$
in the zeroth--order Hamiltonian $H_0$.
This $E_{\rm imp}(r)$ is constructed in the following way:
We divide the range of $r$ into three regions by introducing
ultraviolet (UV) and infrared (IR) distance scales,
$r_{\rm UV}$ and $r_{\rm IR}$, see Fig.~\ref{potential}.
\begin{itemize}
\item[(i)]
At intermediate distances, $r_{\rm UV} < r < r_{\rm IR}$, where
the bulk of the bottomonium wave functions are located,
$E_{\rm imp}(r)$ is identified with the total energy of the
static $b\bar{b}$ system
$E^{b\bar{b}}_{\rm tot}(r) |_{\mu = \mu_2(r)}
   = [2 m_{b}+V_{\rm QCD}(r)]|_{\mu = \mu_2(r)}$
computed in \cite{RS1}.
$E^{b\bar{b}}_{\rm tot}(r)$ 
is defined in Eq.~(5) of that paper and depends on the parameters 
$\alpha_S^{(3)}(\mu)$, $\mu$, $\overline{m}_b$ and $\overline{m}_c$:
these are, respectively, 
the strong coupling constant defined
in the $\overline{\rm MS}$ scheme with 3 active flavours,
the renormalization scale,
the $b$--quark $\overline{\rm MS}$ mass renormalized at the 
$b$--quark $\overline{\rm MS}$ mass
scale and the same for the $c$--quark.
The $b$--quark pole mass $m_b$ is reexpressed in terms of 
$\overline{m}_b$, such that the ${\cal O}(\Lambda_{\rm QCD})$
renormalons are cancelled in $E^{b\bar{b}}_{\rm tot}(r)$.
In identifying $E_{\rm imp}(r)$ with $E^{b\bar{b}}_{\rm tot}(r)$,
the scale $\mu$ is determined as a function of $r$
according to the second prescription 
($\mu=\mu_2(r)$, Eq.~(14)) in \cite{RS1}:
since $E^{b\bar{b}}_{\rm tot}(r)$ is less $\mu$--dependent and
its series expansion converges better 
if we choose a larger value for $\mu$ when $r$ is smaller, and 
if we choose a smaller value for $\mu$ when $r$ is larger, we consider
our choice of $\mu$ to give a more accurate prediction for
$E^{b\bar{b}}_{\rm tot}(r)$ than
choosing some fixed ($r$--independent) value of $\mu$; 
see Figs.~\ref{scales},\ref{mudep}.
\begin{figure}
\begin{center}
\psfrag{R}{$r\,[{\rm GeV}^{-1}]$}
\psfrag{Energy}{$E^{b\bar{b}}_{\rm tot}$ [GeV]}
\psfrag{n=1}{${\cal O}(\alpha_s^1)$}
\psfrag{n=2}{${\cal O}(\alpha_s^2)$}
\psfrag{n=3}{${\cal O}(\alpha_s^3)$}
\psfrag{mu=1}{$\mu=1$~GeV}\psfrag{a}{a)}
\includegraphics[width=7cm]{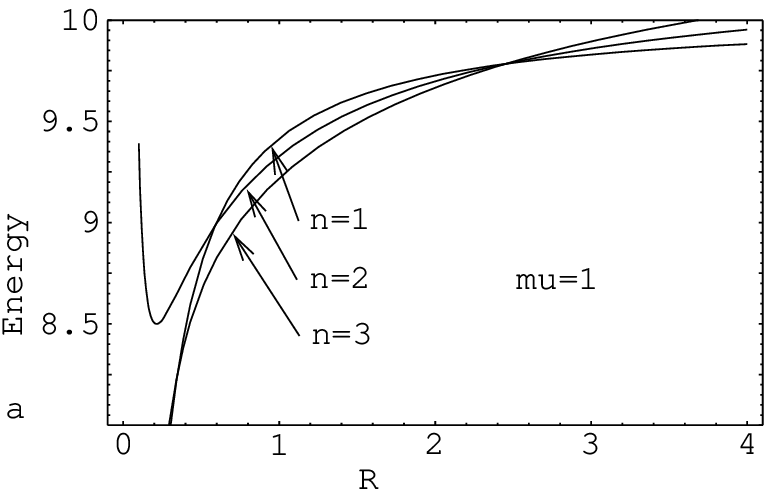}
\psfrag{R}{$r\,[{\rm GeV}^{-1}]$}\psfrag{Energy}{}
\psfrag{mu=2}{$\mu=2$~GeV}\psfrag{b}{b)}
\includegraphics[width=7cm]{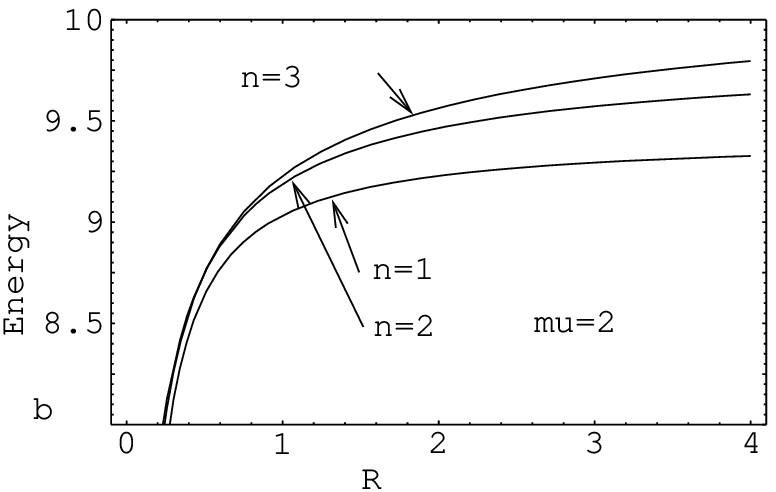}
\end{center}
\vspace*{-.5cm}
\caption{\small 
$E^{b\bar{b}}_{\rm tot}(r) = 2 m_{b}+V_{\rm QCD}(r)$ 
up to ${\cal O}(\alpha_S^N)$ for $N=1$, 2, 3
when $\mu$ is fixed independently of $r$:
a) $\mu=1$~GeV, and b) $\mu=2$~GeV.
The scale $\mu$ that provides the best convergence of the
perturbative series depends on $r$.\label{scales}}
%
\begin{center}
\psfrag{mu}{$\mu\,$[GeV]}
\psfrag{Etot}{$E^{b\bar{b}}_{\rm tot}$ [GeV]}
\psfrag{r1}{$r=1 \,{\rm GeV}^{-1}$}
\psfrag{r2}{$r=2 \,{\rm GeV}^{-1}$}
\psfrag{r3}{$r=3 \,{\rm GeV}^{-1}$}
\includegraphics[width=7cm]{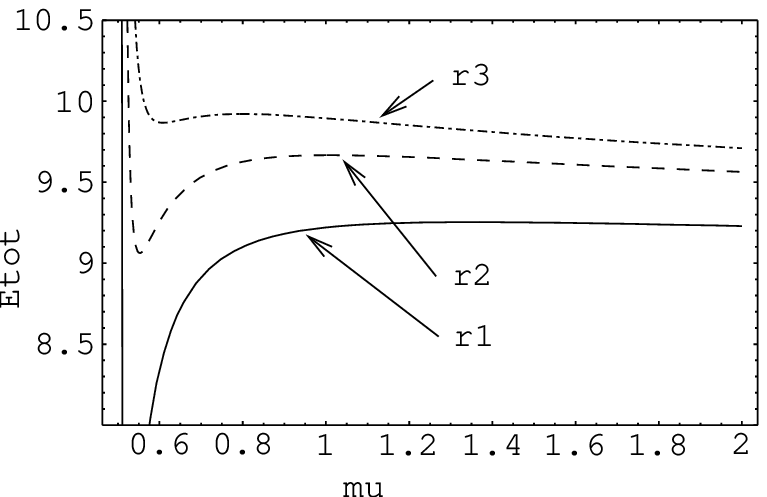}
\end{center}
\vspace*{-.5cm}
\caption{Dependence of $E_{\rm tot}^{b\bar{b}}(r)$ on
the scale $\mu$
for three different values of $r=1,2,3 \,{\rm GeV}$
(solid, dashed, dash--dotted, respectively). 
For large $r$, the flat region (less $\mu$--dependent region) moves
to smaller scales.
\label{mudep}
}
\end{figure}
The $c$--quark $\overline{\rm MS}$ mass is taken as
$\overline{m}_c = 1.243$~GeV \cite{bsv1}.
We will explain how we fix $\overline{m}_b$ in our analysis below.
For other details, we follow the convention of Secs.~IIA and IIB of
\cite{RS1}.

\item[(ii)]
At short distances, $r < r_{\rm UV}$, we use a renormalization--group
improved QCD potential.
It is obtained by integrating the three--loop renormalization--group improved
interquark force $F(r) \equiv - dV_{\rm QCD}(r)/dr$,
following the method of Sec.~4 of \cite{paper1}.
There, it was shown that the QCD
potential becomes more convergent if we improve the interquark
force by means of the renormalization group and integrate it over $r$, 
rather than directly improving the QCD potential by the renormalization 
group.\footnote{
Here, the renormalization-group improvement of $V_{\rm QCD}(r)$
refers to that using the $V$-scheme beta function 
(the second paper of \cite{ps}) and is different from the
renormalization-group improvement in \cite{ams}--\cite{p02}
(resummation of the next-to-next-to-leading logarithms).
}
The initial value for the renormalization group evolution 
of $F(r)$ and the
constant part of $E_{\rm imp}(r)$ are determined such that 
$E_{\rm imp}(r)$ becomes continuous at $r = r_{\rm UV}$
up to the first derivative.

\item[(iii)]
At long distances, $r > r_{\rm IR}$, we use a linear potential,
$E_{\rm imp}(r)=C_1 \, r + C_2$,
where $C_1$ and $C_2$ are determined such that 
$E_{\rm imp}(r)$ becomes continuous at $r = r_{\rm IR}$
up to the first derivative.

\end{itemize}

The main part of the improved potential $E_{\rm imp}(r)$
is that given in the intermediate--distance region (i).
Indeed this part of the potential
dictates the main features of the results of
our analysis.
According to its construction, however, $E_{\rm imp}(r)$ as defined in 
the region (i) becomes unstable and
unreliable at $r \simlt 1/m_b$ and $r \simgt 1/\Lambda_{\rm QCD}$.
This is the reason why we continue it to other definitions
at short and long distances.
The short-- and intermediate--distance parts (i) and 
(ii) are determined by perturbative QCD,
whereas the long--distance part (iii) is not.
We will show that the shape of $E_{\rm imp}(r)$ in the
long--distance region (iii) does not affect
the results of our analysis significantly.
It should be noted that, 
with our definition, the $\mu$--dependence of $E_{\rm imp}(r)$ has
been completely eliminated. 
The $\mu$-dependences mentioned later
in this article are those which stem from the
other terms of the Hamiltonian (see below).
We urge the reader to consult \cite{paper1,RS1} for the physics background
and detailed features of $E_{\rm imp}(r)$ as defined in 
the regions (i) and (ii).
In our analysis below, we set $r_{\rm UV}=0.5~{\rm GeV}^{-1}$
and $r_{\rm IR}=4.5~{\rm GeV}^{-1}$.

\subsection{Perturbative Expansion}

We solve the non-relativistic Schr\"odinger equation numerically with the
improved potential,
\bea
H_0^{\rm (imp)} \ket{\psi} = E_\psi^{(0)} \ket{\psi},
\qquad
H_0^{\rm (imp)}=\frac{\vec{p}\,^2}{ m_b} + E_{\rm imp}(r) ,
\label{schroedinger-eq}
\eea
and determine the zeroth--order energy level $E_\psi^{(0)}$ 
and wave
function $\ket{\psi}$ of a quarkonium state.
We treat $U$, $W_A$ and $W_{NA}$ as perturbations and calculate
the perturbative corrections to the energy level by
\bea
&&
\Delta E_\psi = \bra{\psi} ( U + W_A + W_{NA} ) \ket{\psi} .
\label{DeltaE}
\eea
We will also examine corrections induced by
some of the ${\cal O}(1/c^3)$ operators:
\bea
\delta U &=& \delta U_{LS} \, \, \vec{L}\cdot\vec{S}  
+ \delta U_{S}  \, \,
\biggl[ {S^2} - 3 \frac{\left(\vec{S}\cdot\vec{r}\right)^2}{r^2} \biggr] ,
\\ 
\delta U_{LS} &=&
\frac{3 C_F \alpha_S^{(n_l)}}{2 m_b^2 r^3} \, 
\times \,
\frac{\alpha_S^{(n_l)}}{\pi}
\left\{
\frac{\beta_0}{2} \left( \ell_\mu -1 \right) 
- \frac{2}{3}\, C_A \left( \ell_m - \frac{55}{24} \right)
+ \frac{2}{3}\, C_F - \frac{5}{9} \, T_R n_l  
\right\} ,
\\ 
\delta U_{S} &=&
- \frac{C_F \alpha_S^{(n_l)}}{2 m_b^2 r^3} \, 
\times
\frac{\alpha_S^{(n_l)}}{\pi}
\left\{
\frac{\beta_0}{2} \left( \ell_\mu - {4\over3} \right) 
- C_A \left( \ell_m - {97\over36} \right)
+ C_F - \frac{5}{9}\, T_R n_l 
\right\} ,
\\ 
\delta W_{NA} &=& \frac{C_F (\alpha_S^{(n_l)})^2}{m_b r^2} \, 
\times \,
\frac{\alpha_S^{(n_l)}}{\pi}
\left\{ \frac{\left(C_F-2C_A\right)\beta_0}{4}\, 
\ell_\mu + {b_2\over 2}\right\}
\nonumber \\ &&
  \qquad
- \frac{(C_F \alpha_S^{(n_l)})^2}{4 m_b r^2} \, 
\times \,
\frac{\alpha_S^{(n_l)}}{\pi}
\left\{ {\beta_0}\left(\ell_\mu - \frac{3}{4}\right)+\frac{a_1}{4}\right\}
,
\label{deltaWNA}
\eea
where
\bea
\ell_\mu &=& \log (\mu\, r) + \gamma_E ,
\\
\ell_m &=& \log (m_b\, r) + \gamma_E ,
\\
\beta_0 &=& \frac{11}{3}\, C_A - \frac{4}{3}\, T_R n_l ,
\\
a_1 &=& {31\over 9}\, C_A -{20\over 9}\, T_R n_l ,
\\
b_2 &=& \left({65\over 18}-{8\over 3}\log 2\right)C_F C_A
-{2\over 9}\, C_F T_R n_l
\nonumber \\ &&  
-\left({101\over 36}+{4\over 3}\log 2\right)C_A^2+{49\over 36}\, C_A T_R n_l
.
\eea
$\gamma_E=0.5771...$ is the Euler constant; 
$T_R = 1/2$.
The spin--dependent operators
$\delta U_{LS}$ and $\delta U_S$ were derived in \cite{manstew,kpss2}.
(Earlier incomplete results can be found in \cite{grr,yn3}.)
We derived $\delta W_{NA}$ from the result of \cite{kpss1} 
in the following way:
We discard the logarithm originating from the
IR divergence associated with the energy scale.
This gives the first term of Eq.~(\ref{deltaWNA}).
We have to take into account the contributions which come from
the unitary transformation, 
$H = e^X H' e^{-X}$, where $H'$ is the Hamiltonian of \cite{kpss2}
and\footnote{
The operator representation of $X$ reads
$$
X = - \frac{C_F \alpha_S^{(n_l)}}{4 m_b r} \, 
\left[
\Bigl( 1 + i \vec{r}\cdot\vec{p} \Bigr)
+ \frac{\beta_0\alpha_S^{(n_l)}}{4\pi}
\Bigl\{
2 \ell_\mu + ( 2 \ell_\mu - 1 ) i \vec{r}\cdot\vec{p}
\Bigr\}
\right] + {\cal O}(\alpha_S^3) .
$$
} 
\bea
\bra{\vec{p}+\vec{k}} X \ket{\vec{p} \rule{0mm}{4mm}} = 
- \frac{\pi C_F \alpha_S^{(n_l)}(k)}{m_b k^2}
\left( 1 + \frac{2 \vec{p}\cdot\vec{k}}{k^2} \right) ,
 \qquad
k = |\vec{k}| .
\eea
The second term of Eq.~(\ref{deltaWNA}) is generated by
this unitary transformation.
We note that the non--logarithmic part of $\delta W_{NA}$ cannot be
determined unambiguously, 
since it mixes with other ${\cal O}(1/c^3)$ operators 
through IR divergence.
Our definition merely represents one possible scheme.
Only when we add all the contributions to the spectrum
at ${\cal O}(1/c^3)$, the sum is free from IR divergence
and can be defined unambiguously.
Since at present we do not know the full form of the Hamiltonian
up to ${\cal O}(1/c^3)$,\footnote{
The ${\cal O}(1/c^3)={\cal O}(\alpha_S^4)$ non-logarithmic term of
$V_{\rm QCD}(r)$ is not known yet, which would mix with $W_{NA}$
through IR divergence.
All other ${\cal O}(1/c^3)$ operators of
the Hamiltonian have been identified in Ref.~\cite{kpss2}.
}
this problem cannot be circumvented in any case.

We have to specify how we treat the pole mass $m_b$ in the operators
$\vec{p}\,^2/m_b$, $U$, $W_A$, $W_{NA}$, $\delta U$ and $\delta W_{NA}$.
We express the pole mass $m_b$ in terms of the $\overline{\rm MS}$ mass
$\overline{m}_b$ in the series expansion in $\alpha_S^{(4)}(\mu)$
up to ${\cal O}(\alpha_S^3)$
using Eqs.~(2), (7) and (8) of \cite{RS1}.\footnote{
These formulas were derived originally in \cite{chst,mr,hoangmceff}.
}
After that we can, in principle, reexpand these operators in
$\alpha_S^{(4)}(\mu)$, since the pole mass enters the denominators
and $\ell_m$.
There is, however, no known guiding principle how to organize these expansions,
since such reexpansions cannot be carried out consistently with the
expansions in $1/c$.\footnote{
For instance, if we organize the perturbative expansions of the energy
levels appropriately, renormalons contained in them are of order 
$\Lambda_{\rm QCD}^4$, 
whereas if one expands the levels in $1/c$, renormalons become 
order $\Lambda_{\rm QCD}^3$ and worsen the perturbative convergence 
\cite{offshell}.  
This may be regarded as an explicit example of problems due to an
inconsistency between $1/c$ power counting and renormalon cancellations
(beyond those controlled by $\epsilon$--expansion \cite{hlm}).  
}
This is in contrast to the reexpansion of 
$E^{b\bar{b}}_{\rm tot}(r) = 2 m_{b}+V_{\rm QCD}(r)$, 
which has a guiding principle by the cancellation of 
${\cal O}(\Lambda_{\rm QCD})$ renormalons, although the
reexpansion is indeed inconsistent with the $1/c$--expansion.
Hence, we keep the pole mass as a function of
$\overline{m}_b$, $\alpha_S^{(4)}(\mu)$ and $\mu$ and do not
reexpand the operators 
($\vec{p}\,^2/m_b$, $U$, $W_A$, $W_{NA}$, $\delta U$, $\delta W_{NA}$);
the values of the pole mass are shown in Tab.~\ref{polemass} corresponding to
$\alpha_S^{(5)}(M_Z)=0.1181$ and $\overline{m}_b=4.190$~GeV.
We will examine how the uncertainty of the pole mass of
order $\Lambda_{\rm QCD}$ affects our predictions in Sec.~6.
\begin{table} \begin{center}
\begin{tabular}{|c|c|c|c|c|c|}
\hline
\parbox{5mm}{$\mu$} [GeV] & 1.0 & 2.0 & 3.0 & 4.0 & 5.0 \\
\hline
\parbox{5mm}{$m_b$} [GeV] & 5.458 & 5.131 & 5.027 & 4.969 & 4.930 \\
\hline
\end{tabular}
\caption{\small
Depencence of the pole mass 
$m_b(\overline{m}_b,\alpha_S^{(4)}(\mu),\mu)$ on the scale $\mu$ for
$\alpha_S^{(5)}(M_Z)=0.1181$ and $\overline{m}_b=4.190$~GeV.\label{polemass}}
\end{center} \end{table}

We take the input for the strong coupling constant as $\alpha_S^{(5)}(M_Z)$
and calculate $\alpha_S^{(3)}(\mu_2(r))$ for $E_{\rm imp}(r)$ and
$\alpha_S^{(4)}(\mu)$ for the other operators.
We evolve the coupling by solving
the 3--loop renormalization-group equation numerically and match it
to the 4-- and 3--flavour couplings successively through the matching
condition \cite{lrv}.\footnote{
The matching scales are taken as $\overline{m}_b$ and $\overline{m}_c$,
respectively.
}
(Although the 4--loop running of the $\overline{\rm MS}$ coupling
constant is available, we consider the 3--loop running more
consistent in our analysis, which incorporates corrections up to the
2--loop finite part of $V_{\rm QCD}(r)$.)
\medbreak

Our choice of the zeroth--order Hamiltonian and the way we organize the
perturbative expansion is
largely motivated by the success of phenomenological potential models.
In fact, the above organization of the perturbative
expansion follows, to a large extent, the approaches of phenomenological model
analyses, if we identify our $H_0^{\rm (imp)}$ 
with the non--relativistic Hamiltonian, 
$p^2/m + V_{\rm pheno}(r)$, used in those analyses.
Since $E_{\rm imp}(r)$ agrees well with typical phenomenological
potentials up to an additive constant, we expect that we can make
close comparisons with phenomenological model analyses, and that
eventually the bottomonium spectrum may be reproduced with a good accuracy.
Note that our zeroth--order quarkonium wave functions, which are
determined from $H_0^{\rm (imp)}$, include some of the higher--order corrections in the usual
order counting of the $1/c$--expansion, since $E_{\rm imp}(r)$ includes the
${\cal O}(1/c)$ and ${\cal O}(1/c^2)$ corrections to the
static QCD potential.
By the same token, the zeroth--order energy levels are different from the
Coulomb energy levels, and in particular they depend on the
orbital angular momentum $l$.

\section{Zeroth--order Energy Levels, Wave Functions and Scales}
\label{s3}

\begin{figure}
\begin{center}
\psfrag{r}{$r\,[{\rm GeV}^{-1}]$}
\psfrag{rPsi2}{\hspace{-6mm}\footnotesize $r^2 |R_\psi(r)|^2$ [GeV]}
\psfrag{1S}{$1S$}
\includegraphics[width=5.4cm]{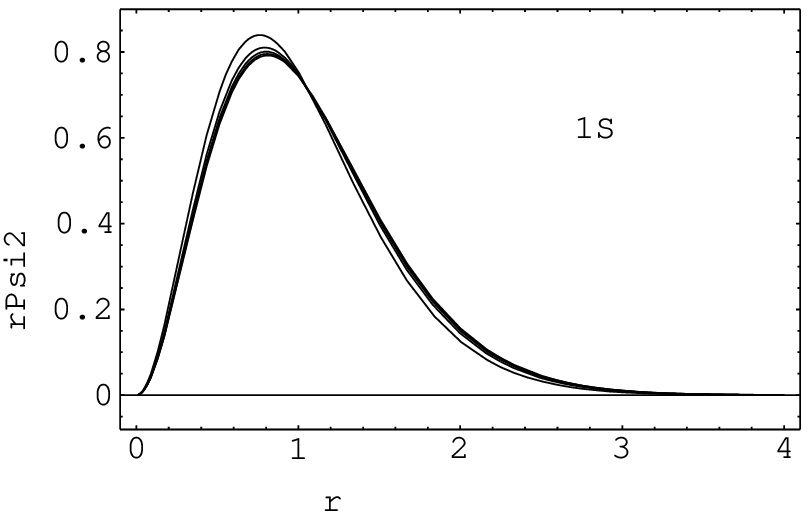}
\hspace{-.5cm}
\psfrag{rPsi2}{}\psfrag{2S}{$2S$}
\includegraphics[width=5.4cm]{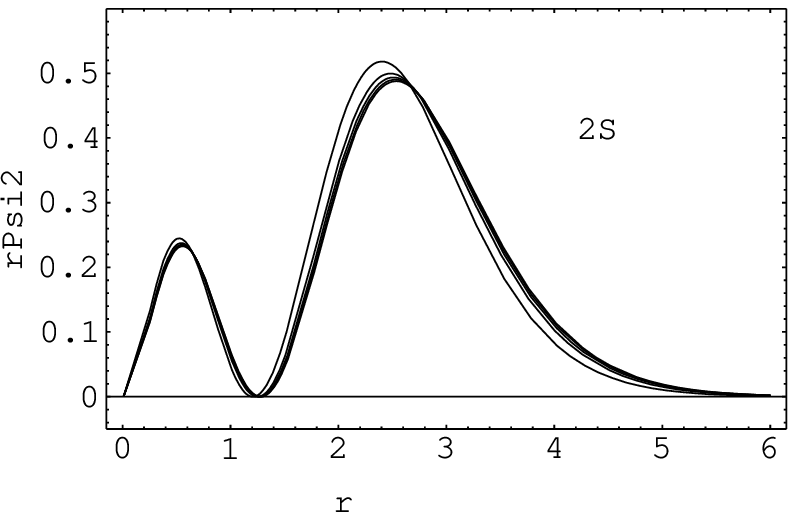}
\hspace{-.5cm}
\psfrag{3S}{$3S$}
\includegraphics[width=5.4cm]{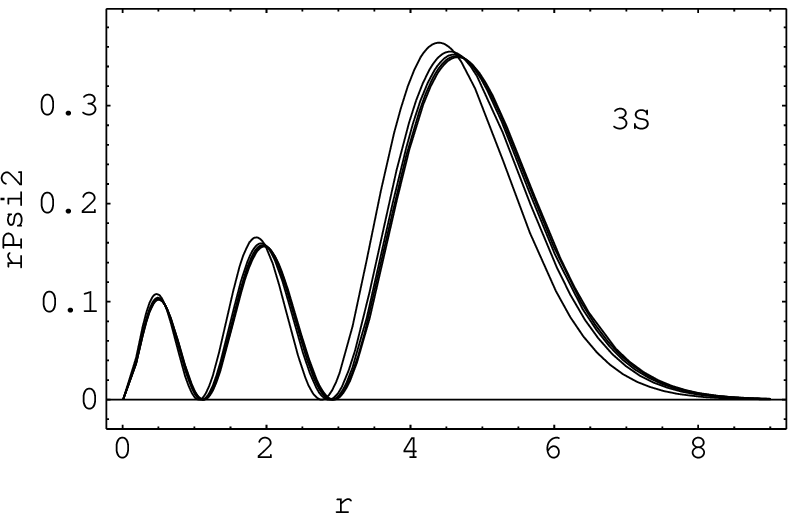}
\end{center}
\vspace*{-.5cm}
\begin{center}
\psfrag{r}{$r\,[{\rm GeV}^{-1}]$}
\psfrag{rPsi2}{\hspace{-6mm}\footnotesize $r^2 |R_\psi(r)|^2$ [GeV]}
\psfrag{1P}{$1P$}
\includegraphics[width=5.4cm]{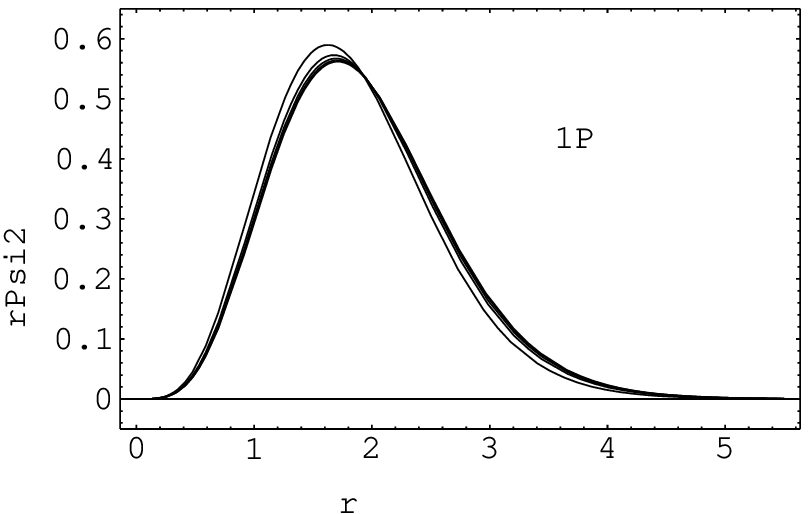}
\hspace{-.5cm}
\psfrag{rPsi2}{}\psfrag{2P}{$2P$}
\includegraphics[width=5.4cm]{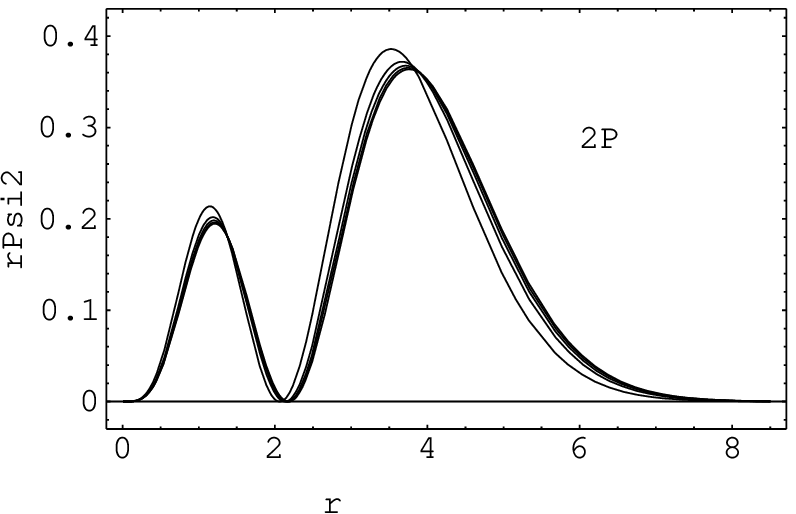}
\end{center}
\vspace*{-.5cm}
\caption{\small
$r^2 |R_\psi(r)|^2$ for different values of $\mu=1,2,3,4$ and $5$ GeV,
where $R_\psi(r)$ denotes the radial part of the zeroth-order
energy eigenfunction.
The area below each curve is normalized to unity.
The one curve that visibly differs from the others corresponds to $\mu=1$~GeV.
Note that the scales differ between the plots.
\label{wavefunctionsmu}}
\end{figure}

We show numerical solutions to the zeroth--order
Schr\"odinger equation (\ref{schroedinger-eq}).
Here and hereafter we take the input value for the strong coupling
constant to be the present world average value
$\alpha_S^{(5)}(M_Z)=0.1181$ \cite{pdg}.
Through Secs.~\ref{s3}--\ref{s5}
we use the bottom quark $\overline{\rm MS}$ mass 
$\overline{m}_b=4.190$~GeV taken from \cite{bsv2}.
We show the zeroth--order energy levels $E_\psi^{(0)}$ in 
Tab.~\ref{zerothorderenergy}.
\begin{table}
\begin{center}
\begin{tabular}{|c|c|c|c|c|c|}
\hline 
$\mu\,$ & $1S$ & $1P_j$ & $2S$ & $2P_j$ & $3S$\\
\hline 
1.0 & 9.476 & 9.877 &  9.986 & 10.186 & 10.247\\
2.0 & 9.498 & 9.896 & 10.007 & 10.203 & 10.262\\
3.0 & 9.505 & 9.902 & 10.013 & 10.209 & 10.268\\
4.0 & 9.509 & 9.906 & 10.017 & 10.212 & 10.271\\
5.0 & 9.512 & 9.908 & 10.020 & 10.214 & 10.272\\
\hline
\end{tabular}
\caption{\small Zeroth--order energies $E_\psi^{(0)}$. 
The units are GeV.
Unless otherwise
stated, all tables and figures use $\overline{m}_b=4.190$~GeV,
$\alpha_S^{(5)}(M_Z)=0.1181$ and the pole mass
$m_b(\overline{m}_b,\alpha_S^{(4)}(\mu),\mu)$ as explained in the text.
\label{zerothorderenergy}}
\end{center}
\end{table}
The squared radial wave functions multiplied by the phase space
factor are shown in Fig.~\ref{wavefunctionsmu}.
Both the energy levels and the wave functions
depend on the scale $\mu$ (only) through the pole mass 
in $\vec{p}\,^2/m_b$ in the Schr\"odinger equation.
The energy levels $E_\psi^{(0)}$ vary by about 10~MeV (20~MeV) when 
$\mu$ is varied from 2 to 5~GeV (1 to 2~GeV).
Nonetheless, if we take the difference of any of the two energy levels,
the $\mu$ dependences cancel mostly.
The $\mu$-dependences of the wave functions are fairly weak.

\begin{figure}
\begin{center}
\psfrag{q}{$q\,[{\rm GeV}]$}\psfrag{fx}{}
\psfrag{f1S}{$f_{1S}$}\psfrag{f2S}{$f_{2S}$}\psfrag{f3S}{$f_{3S}$}
\psfrag{mu1S}{$\mu_{1S}$}\psfrag{mu2S}{$\mu_{2S}$}\psfrag{mu3S}{$\mu_{3S}$}
\psfrag{alphas}{$\alpha_s$}
\includegraphics[width=12cm]{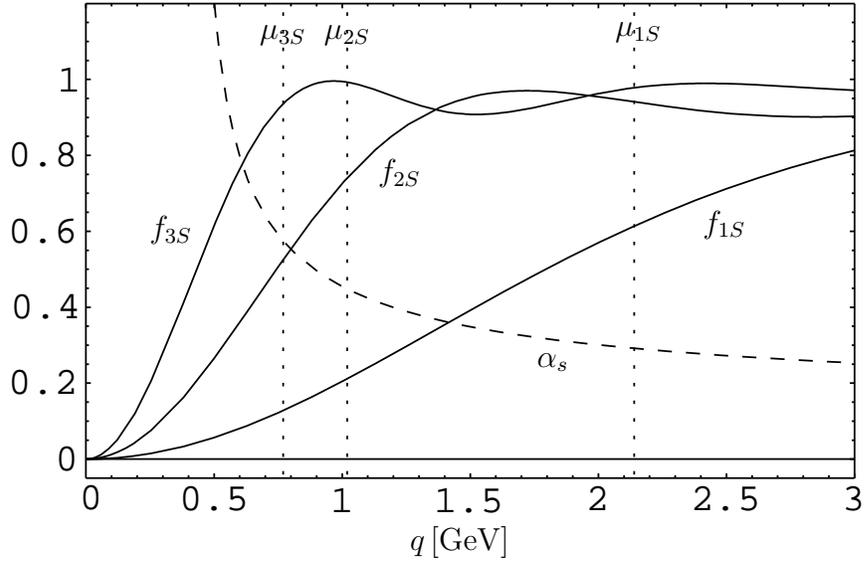}
\end{center}
\vspace*{-.5cm}
\caption{\label{supportS}
\small 
Support functions for the $S$ states.
The solid curves show the support functions as defined in
Eq.~(\ref{supportfunctiondef}); for comparison of the relevant
scales, $\alpha_s^{(4)}(\mu)$ is also plotted (dashed curve). Since
the analysis that we advocate in this work does not attribute
scales to the individual states, the scales indicated by the
dotted lines are taken from \cite{bsv2}, Table~II.}
\end{figure}

\begin{figure}
\begin{center}
\psfrag{q}{$q\,[{\rm GeV}]$}\psfrag{fx}{}
\psfrag{alphas}{$\alpha_s$}
\psfrag{f1P}{$f_{1P}$}\psfrag{f2P}{$f_{2P}$}
\psfrag{mu1P}{$\mu_{1P}$}\psfrag{mu2P}{$\mu_{2P}$}
\includegraphics[width=12cm]{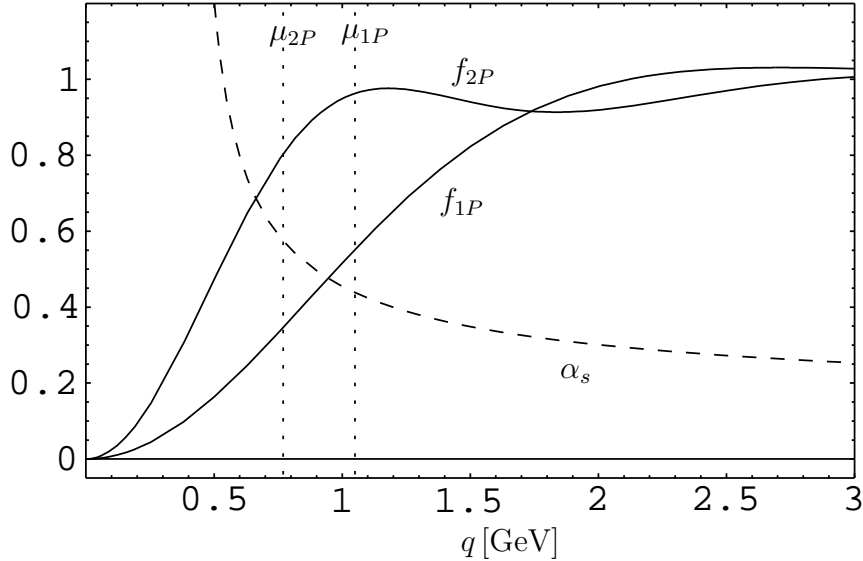}
\end{center}
\vspace*{-.5cm}
\caption{\small Support functions for the $P$ states.
Notations are same as in Fig.~\ref{supportS}.
\label{supportP}}
\end{figure}

In \cite{bsv1,bsv2} the scale $\mu=\mu_\psi$ for each quarkonium state 
$\ket{\psi}$ was fixed
by minimizing the $\mu$-dependence of each
energy level calculated in a fixed--order perturbative expansion
(minimal--sensitivity prescription).
The scale fixed in this way turns out to
represent the physical size of the
corresponding quarkonium state fairly well.
This was shown in \cite{bsv1} by comparing the scale $\mu_\psi$ and 
the support function defined by
\bea \label{supportfunctiondef}
f_\psi(q) = \theta (\overline{m}-q) 
- \int_0^\infty dr \, r^2 \, \left| R_\psi(r) \right|^2 \, \frac{\sin (qr)}{qr},
\eea
where $R_{\psi}(r)$ is the radial part of the wave function for
the state $\ket{\psi}$.
The support function represents the support in the momentum--space 
integral in the calculation of (the major part of) the energy level.
The Coulomb wave function, evaluated with
$\alpha_S^{(4)}(\mu_\psi)$, was used to compute the support function
$f_\psi(q)$; the charm mass effects in loops were not taken into account.
Here we compare the support function computed
with our zeroth--order wave function
and the scale fixed in \cite{bsv2}, both of
which include the charm mass effects. The resulting support functions
for the $S$ states are shown in Fig.~\ref{supportS}
and for the $P$ states in Fig.~\ref{supportP}.
We see that with respect to the treatment in \cite{bsv1}, the support
functions are shifted slightly towards higher momentum.
This is reasonable, since the wave functions calculated here are
peaked closer to the origin in coordinate space than the Coulomb
wave functions (see e.g.\ Fig.~\ref{wavefunctions} below).
Since the scales $\mu_\psi$ are located within the IR ``tails'' of the
corresponding support functions $f_\psi(q)$, we confirm that the
above interpretaion of the minimal-sensitivity scale
$\mu_\psi$ is valid also when we include the
charm mass effects and the ${\cal O}(1/c)$ and ${\cal O}(1/c^2)$ 
corrections of $V_{\rm QCD}(r)$ to the wave functions.

\section{Fine Splittings}
\clfn

In this section we examine the fine splittings in the bottomonium
spectrum within our framework and compare them with the experimental data
as well as with previous theoretical predictions in the literature.
Within perturbative QCD, it is expected that in principle
the fine splittings can be computed
much more accurately than the individual energy levels.
This is because the potentials which contain order
$\Lambda_{\rm QCD}^3 r^2$ renormalons 
(the static QCD potential and $W_{NA}$)
do not induce the fine splittings,
so that these renormalons cancel in the computation of the fine
splittings.

The fine splittings (or fine structure) are the level splittings
among the states 
with the same principal quantum number $n$, 
orbital--angular momentum $l>0$ and spin $s$ but
with different total angular momentum $j$.
Experimentally, the fine splittings have been observed among the
$1^3P_j$ [$(n,l,s)=(2,1,1)$] and $2^3P_j$ [$(n,l,s)=(3,1,1)$] states,
we therefore examine the predictions for these splittings.

In perturbative QCD the fine splittings are induced first at 
${\cal O}(\alpha_S^4 m_b) = {\cal O}(1/c^2)$
by the operator $U$:
\bea
\Delta E_U^{\rm (C)} =
\bra{\psi_{\rm C}} U \ket{\psi_{\rm C}} &=& 
\frac{(C_F \alpha_S^{(n_l)})^4}{8\, n^3} \, m_b \times
{\frac{D_S + 3 X_{LS}  }{l\,
      \left( l+1 \right) \,\left( 2\,l +1 \right) }} 
\qquad
(l > 0) ,
\label{fsinC}
\eea
where
\bea
D_{S} &\equiv&
- \left< 
\vec{S}^2 - 3 \frac{(\vec{r}\cdot \vec{S})^2}{r^2} 
\right>
=
\frac{
2 l (l+1) s (s+1) - 3 X_{LS} - 6 X_{LS}^2
}{
(2l-1)(2l+3)
},
\\
X_{LS} &\equiv&
\left< \vec{L}\cdot \vec{S} \right>
= \frac{1}{2}\,
\left[ j(j+1)-l(l+1)-s(s+1) \right] .
\eea
This is the fixed--order formula used in \cite{bsv1,bsv2}.
In this formula, the Coulomb wave function $\ket{\psi_{\rm C}}$
is used to compute the expectation value.
The scale dependence of $\Delta E_U^{\rm (C)}$ is large,
since it is proportional to $[\alpha_S^{(4)}(\mu)]^4$:
three powers of $\alpha_S$ come from the Coulomb wave function
$\bra{\psi_{\rm C}} \times \ket{\psi_{\rm C}}$, and one power comes from
the operator $U$.
In \cite{bsv1,bsv2} the scale $\mu$
is fixed by the minimal--sensitivity prescription; c.f.\ the previous
section.

In our approach, we calculate the fine splittings from
$\Delta E_U = \bra{\psi} (U+\delta U) \ket{\psi}$.
As compared to the fixed--order formula, some of the higher--order
corrections are incorporated through the wave function $\ket{\psi}$
and the operator $\delta U$.

In phenomenological approaches, one computes the fine splittings
using the same operator $U$ but using the wave functions
determined from phenomenological potentials:
\bea
\Delta E_U^{\rm (pheno)} = \bra{\psi_{\rm pheno}} U \ket{\psi_{\rm pheno}} .
\eea
Some of the higher-order corrections to $U$ constrained by
the Gromes relation \cite{gromes}
have also been incorporated.
As an example we compare our results with those of a phenomenological
model with a Coulomb--plus--linear potential (Cornell potential)
\cite{cornell}:
\bea
&&
H_0^{\rm (Cornell)} = \frac{\vec{p}\,^2}{m_b} + V_{\rm Cornell}(r) ,
\\ &&
V_{\rm Cornell}(r) = - \frac{\kappa}{r} + \frac{r}{a^2}
\eea
with $\kappa = 0.52$ and $a = 2.34$~GeV$^{-1}$.

\begin{table}
\begin{center}
\begin{tabular}{|c|c|c|c|c|c|c|c|}
\hline
  & \multirow{2}{.4cm}{$\alpha_s$} 
  &  \multicolumn{2}{|c|}{$\left<\psi_{\rm C  }\left| \times \right|\psi_{\rm C}\right>$}
  &  \multicolumn{2}{|c|}{$\left<\psi_{\rm QCD}\left| \times \right|\psi_{\rm QCD}\right>$}
  &  \multicolumn{2}{|c|}{$\left<\psi_{\rm Cornell}\left| \times \right|\psi_{\rm Cornell}\right>$}\\
 && \parbox{1cm}{\centering $U_{LS}$} & \parbox{1cm}{\centering $U_S$} & 
    \parbox{1.3cm}{\centering $U_{LS}$} & \parbox{1.3cm}{\centering $U_S$} & 
    \parbox{1.6cm}{\centering $U_{LS}$} & \parbox{1.6cm}{\centering $U_S$}\\
\hline
\multirow{2}{.6cm}{$1P$} & 0.360 & 2.08 & $-$0.69 & 13.71 & $-$4.57 & 16.07 & $-$5.36 \\
                       & 0.468 & 5.95 & $-$1.98 & 17.82 & $-$5.94 & 20.89 & $-$6.96 \\ 
\hline
\multirow{2}{.6cm}{$2P$} & 0.360 & 0.62 & $-$0.21 &  8.19 & $-$2.73 & 12.92 & $-$4.31 \\
                       & 0.726 & 10.22 & $-$3.41 & 16.52 & $-$5.51 & 26.06 & $-$8.69 \\ 
\hline
\end{tabular}
\caption{\small Expectation values (in MeV) of the operators $U_{LS}$ and $U_S$. 
The mass used in the operators is $m_b=5.027$~GeV corresponding 
to $\overline{m}_b=4.190$~GeV and $\mu=3$~GeV. \label{UlsandUs}}
\end{center}
\end{table}
Tab.~\ref{UlsandUs} compares the expectation 
values of the operators $U_{LS}$ and $U_S$
with respect to the Coulomb wave function $\ket{\psi_{\rm C}}$, 
our zeroth--order wave function $\ket{\psi}$ and
the Cornell wave function $\ket{\psi_{\rm Cornell}}$. 
The operators $U_{LS}$
and $U_S$ induce the fine splittings through
$X_{LS} \,U_{LS} - D_S \,U_S$, where $(X_{LS},D_S)=
(-2,-2),(-1,1),(1,-1/5)$ for $j=0,1,2$, respectively, and 
$(l,s)=(1,1)$.
The fine splittings between adjacent levels are therefore
$U_{LS} - 3 \,U_S$ for $P_1-P_0$ and $2 \,U_{LS} + 6\,  U_S/5$ 
for $P_2-P_1$.
The values of $\alpha_S^{(4)}(\mu)$ in the operators $U_{LS}$ and $U_S$
are taken as
0.36 and 0.468 for the $1P_j$ states and as
0.36 and 0.726 for the $2P_j$ states.
Also the same values of $\alpha_S^{(4)}(\mu)$ are used for 
calculating the Coulomb wave functions 
$\ket{\psi_{\rm C}}$.
The first value (0.36) corresponds to the value used in the
phenomenological analysis \cite{eq}.
The latter values (0.468 and 0.726)
are those for the $1P_1$ and $2P_1$ states
which were determined by the minimal--sensitivity prescription in \cite{bsv2}.
Taking into account the numerical values
from the table, $U_{LS}$ and $U_S$ give roughly the same
contribution to the $P_1$--$P_0$--splitting, while the
$P_2$--$P_1$--splitting is dominated by $U_{LS}$.

\begin{figure}
\begin{center}
\psfrag{R}{$r\,[{\rm GeV}^{-1}]$}
\psfrag{WF}{\hspace{-10mm}$r^2\left|R_\psi(r)\right|^2$~~~[GeV]}
\psfrag{QCD}{QCD wave function (solid)}
\psfrag{Cornell}{Cornell wave function (dashed)}
\psfrag{Coulomb}{Coulomb wave function}
\includegraphics[width=12cm]{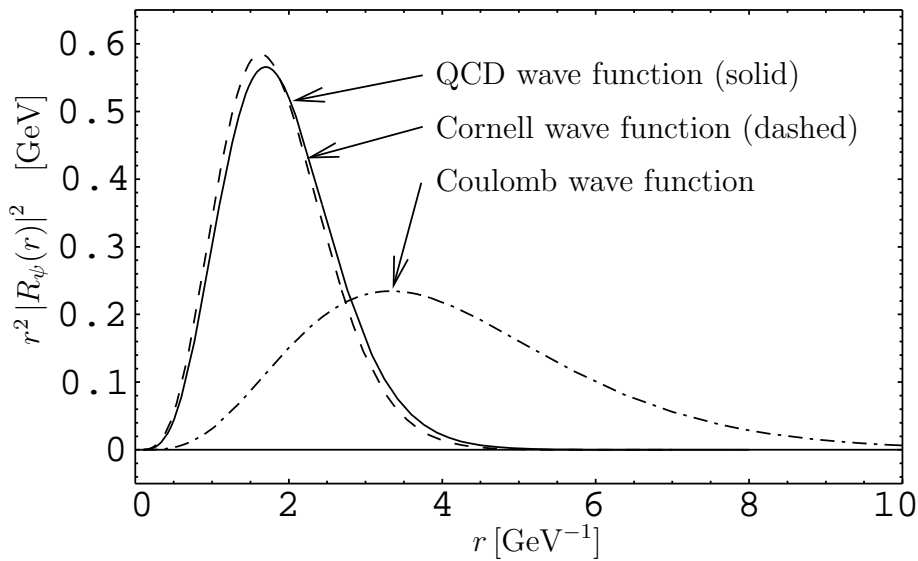}
\end{center}
\vspace*{-.5cm}
\caption{\small Comparison of QCD (solid), Cornell (dashed) and 
Coulomb (dash--dotted) $1P$--wave functions. \label{wavefunctions}
For the Coulomb wave function $\alpha_S=0.36$ has been used.}
\end{figure}

\begin{figure}
\begin{center}
\psfrag{R}{$r\,[{\rm GeV}^{-1}]$}
\psfrag{WFU}{\hspace{-10mm}$r^2\left|R_\psi(r)\right|^2 U_{LS}$~~~[GeV$^2$]}
\psfrag{QCD}{QCD wave function}
\psfrag{Cornell}{Cornell wave function}
\psfrag{Coulomb}{Coulomb wave function}
\includegraphics[width=13cm]{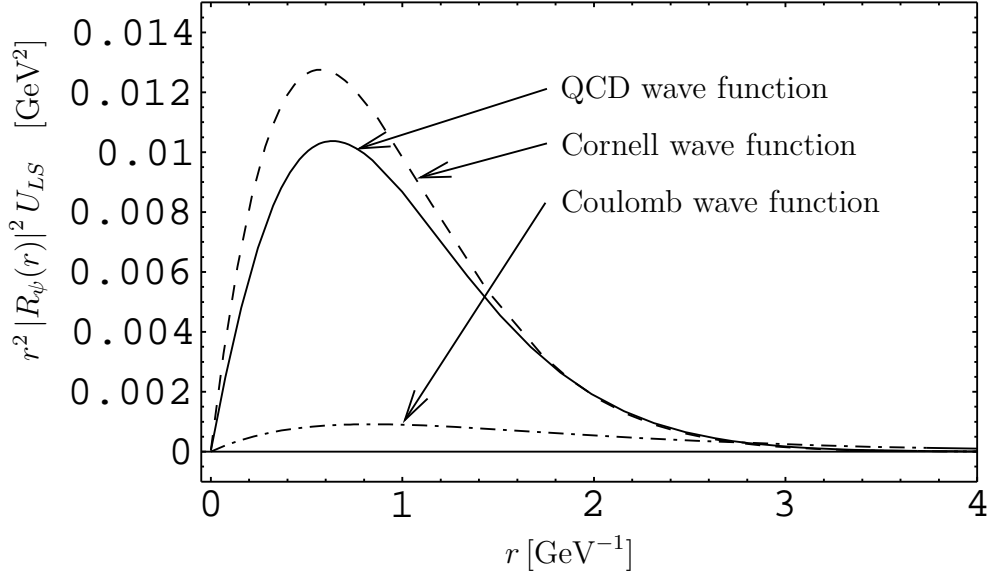}
\end{center}
\vspace*{-.5cm}
\caption{\small Comparison of the integrands 
for $\bra{1P} U_{LS} \ket{1P}$, the conventions are
the same as in Fig.~\ref{wavefunctions}. $\alpha_S=0.36$ has
been used both in the operator $U_{LS}$ and in the determination of 
the Coulomb wave function. Note that the scale of the horizontal
axis is different from Fig.~\ref{wavefunctions}.\label{integrands}}
\end{figure}

We see that the expectation values of $U_{LS}$ and $U_S$
with respect to our wave function
$\ket{\psi}$ and the Cornell wave function $\ket{\psi_{\rm Cornell}}$
are very much larger than the expectation values with respect to
the Coulomb wave function $\ket{\psi_{\rm C}}$.\footnote{
This is not necessarily true for the $2P$ states with
$\alpha_S = 0.726$, but this can be regarded as originating from another
effect, which we explain below.
Namely, the value $\alpha_S = 0.726$ is unrealistically large.
}
The reason for this behaviour can be understood in the following way:
Since the potentials $E_{\rm imp}(r)$ and $V_{\rm Cornell}(r)$
are steeper (i.e.\ the attractive forces are stronger) than the
Coulomb potential in the intermediate--distance region,\footnote{
The cancellation of ${\cal O}(\Lambda_{\rm QCD})$ renormalons
suggests that this behaviour can be understood naturally in terms of
the QCD force
$F(r) = - V_{\rm QCD}'(r) = - C_F \, \alpha_F(1/r)/r^2$ :
$F(r)$ becomes more attractive than the Coulomb force as $r$ increases
due to the running of the $F$--scheme coupling constant $\alpha_F(1/r)$
\cite{paper1,necco-sommer}.
} 
the wave functions are more centered towards the origin
for $\ket{\psi}$ and $\ket{\psi_{\rm Cornell}}$ than for $\ket{\psi_{\rm C}}$;
see Fig.~\ref{wavefunctions}
which compares the squared radial wave functions
multiplied by the phase space factor for the $1P$ states.
Therefore the wave functions 
$\braket{\vec{r}}{\psi}$ and $\braket{\vec{r}}{\psi_{\rm Cornell}}$ 
are enhanced close to
the origin as compared to the Coulomb wave function.
The enhancement factors turn out to be large for the $P$-wave
states.\footnote{
By way of example, if we squeeze the function $r^n$ such that
it takes a same value at half distance, i.e.\
$r^n \to (2r)^n$, then the enhancement factor becomes $2^n$.
Thus, the enhancement factor is larger for a larger $n$.
For the $P$-wave states, $n=4$, because 2 powers come from the wave function
squared and 2 powers come from the phase space.
}
Since the expectation values of $U_{LS}$ and $U_S$ are determined mostly by the
short-distance behaviour of the wave functions, they are enhanced
by large factors.
This feature can be verified in Fig.~\ref{integrands}, 
where we compare the
integrands when the expectation values of $U_{LS}$
are expressed as integrals over $r$;
the main contributions come from distances 
$r \simlt 2~{\rm GeV}^{-1}$.
We should stress that the enhancement of the wave functions
originates mainly from the behaviour of the potentials in 
the intermediate--distance region and not from the short--distance
behaviour.
This can be seen from the fact that
the difference between $\bra{\psi}U_{(LS,S)}\ket{\psi}$ and
$\bra{\psi_{\rm Cornell}}U_{(LS,S)}\ket{\psi_{\rm Cornell}}$ is much 
smaller than the difference between $\bra{\psi}U_{(LS,S)}\ket{\psi}$ and
$\bra{\psi_{\rm C}}U_{(LS,S)}\ket{\psi_{\rm C}}$:
the essential difference between 
our potential $E_{\rm imp}(r)$ and the Cornell 
potential resides in the short--distance region.

We include also the contribution of the ${\cal O}(1/c^3)$ operator
$\delta U$ into our prediction of the fine splittings.
As far as we know, there is no other operator 
which contributes to the fine splittings at ${\cal O}(1/c^3)$.
Considering that our wave function $\ket{\psi}$ includes
all the next--to--leading order [${\cal O}(1/c)$] corrections,
our prediction for the fine splittings incorporates
all the effects up to ${\cal O}(\alpha_S^5 m_b)$
[${\cal O}(1/c)$ relative to the leading ${\cal O}(\alpha_S^4 m_b)$
splittings].
Inclusion of the operator $\delta U$ reduces the scale dependence
of our prediction.
In Tab.~\ref{finesplittings}
\begin{table}
\begin{center}
\begin{tabular}{|c|c|c|c|c|c|c|c|c|c|}
\hline 
\multirow{2}{.3cm}{$\mu$} & \multirow{2}{1cm}{$\alpha_s(\mu)$} &  
 \multicolumn{2}{|c|}{$1P_1-1P_0$} &  \multicolumn{2}{|c|}{$1P_2-1P_1$} &  
 \multicolumn{2}{|c|}{$2P_1-2P_0$} &  \multicolumn{2}{|c|}{$2P_2-2P_1$}\\
&&$U$&$U+\delta U$&$U$&$U+\delta U$&$U$&$U+\delta U$&$U$&$U+\delta U$\\
\hline
1.0 & 0.454 & 33.3 & 22.3 & 26.7 & 16.2 & 20.7 & 12.9 & 16.6 &  9.0\\
2.0 & 0.301 & 22.7 & 24.5 & 18.2 & 18.8 & 13.7 & 14.4 & 11.0 & 10.9\\
3.0 & 0.253 & 19.3 & 23.3 & 15.4 & 18.1 & 11.5 & 13.6 &  9.2 & 10.5\\
4.0 & 0.228 & 17.5 & 22.3 & 14.0 & 17.4 & 10.4 & 13.0 &  8.3 & 10.1\\
5.0 & 0.212 & 16.3 & 21.5 & 13.0 & 16.8 &  9.7 & 12.6 &  7.7 &  9.7\\
\hline
\multicolumn{2}{|c|}{Experiment} & \multicolumn{2}{|c|}{33} &
\multicolumn{2}{|c|}{20} & \multicolumn{2}{|c|}{23} & \multicolumn{2}{|c|}{13}\\
\hline
\end{tabular}
\caption{\small Fine splittings computed from $\bra{\psi}U\ket{\psi}$ and 
$\bra{\psi}U+\delta U\ket{\psi}$. $\mu$ is in GeV, the splittings are
given in MeV.\label{finesplittings}}
\end{center}
\end{table}
we compare the fine splittings calculated from
the matrix elements $\bra{\psi}U\ket{\psi}$
and from $\bra{\psi}U+\delta U\ket{\psi}$.
The former depend on $\mu$ rather strongly and
are larger for smaller $\mu$,
since $U$ is proportional to $\alpha_S^{(4)}(\mu)$.\footnote{
The scale dependence of the wave functions $\ket{\psi}$ through
the pole mass is very weak, c.f.\ Fig.~\ref{wavefunctionsmu}.
}
We see that the scale dependence has decreased considerably
by the effect of $\delta U$.
The scale dependences become minimal at $\mu \simeq 2$~GeV for both
$1P_j$ and $2P_j$ states. 
We may try to reproduce the splittings calculated with
$\bra{\psi}U+\delta U\ket{\psi}$ at these scales
by taking an appropriate choice of the scales in the lower--order
predictions (the splittings calculated with $\bra{\psi}U\ket{\psi}$).
Then the scales become $\mu \simeq 2$~GeV for both
the $1P_j$ and $2P_j$ states.
All these scales are larger than the scales chosen for the
respective states in \cite{bsv2}, 
which are fixed by minimizing the
scale dependence of the individual energy levels.
($\mu \simeq 1$~GeV for the $1P_j$ states and 
$\mu \simeq 0.8$~GeV for the $2P_j$ states; see Fig.~\ref{supportP}.)
This feature is consistent with a naive expectation:
we would expect that the latter scales represent the typical
scales of the binding energies, or the inverse of the sizes of the
boundstates (Fig.~\ref{supportP}), 
whereas the former scales represent those
probed by the operators $U$ or $U + \delta U$, which are 
larger because the contributions to the matrix elements come from
shorter distances (Fig.~\ref{integrands}).

To our knowledge, so far there has been no systematic argument on the
order of renormalons contained in the fine splittings.
Naively we expect that the largest renormalon contained in the
calculation of the fine splittings would be of order
$\Lambda_{\rm QCD}^3/m_b^2 \simeq 1$--10~MeV.
This estimate is based on the Gromes relation \cite{gromes} which tells
us that a part of the
operator $U_{LS}$ is determined from the static QCD potential:
\bea
- \frac{1}{2m_b^2 r}\frac{dV_{\rm QCD}}{dr} \,
\vec{L}\cdot\vec{S}
\eea
and the fact that $dV_{\rm QCD}(r)/dr$ contains an
${\cal O}(\Lambda_{\rm QCD}^3 r)$ renormalon.
We may take this as an order of magnitude estimate of uncertainties of
our present predictions for the fine splittings.
Compared with this error estimate,
our predictions of the fine splittings calculated with
$\bra{\psi}U+\delta U\ket{\psi}$ in Tab.~\ref{finesplittings} are in
reasonable agreement with the experimental data.

We are now able to interpret the reasons why the fine splittings
of the energy levels computed in \cite{bsv2} turned out to be
quite small for the $1P_j$ states but not so much for the $2P_j$ states.
The first point to note is that large
enhancement factors are generated by the fact that the 
quarkonium wave functions
are more centered toward the origin if we solve the 
Schr\"odinger equation with $E_{\rm imp}(r)$
(which we believe to be more realistic) rather than with the Coulomb
potential.
The second point is that the natural scales to be chosen for evaluating
the expectation values of the operator $U$ are larger than
the natural scales for the individual energy levels:
we confirmed this by
incorporating the effects of the higher--order operator $\delta U$,
and the results are qualitatively consistent with a naive expectation.
We find that that the first effect overwhelming the second one
resulted in the quite small splittings among the $1P_j$ states
in \cite{bsv2}.
On the other hand, the cancellation of the first and the second 
effect resulted in reasonable sizes of the fine splittings for
the $2P_j$ states in that paper, which, 
in the light of our present observations, may be regarded 
as rather accidental.

\section{{\boldmath $S$--$P$} Splittings}
\label{s5}

In this section we examine the splittings between the 
$S$--wave and $P$--wave states.
In the Coulomb spectrum the $S$--wave and the $P$--wave states 
with the same principal quantum number $n$ are degenerate.
In perturbative QCD,
the splittings are induced by 
$V_{\rm QCD}(r)$ starting from ${\cal O}(1/c)$ as well as
by $U$, $W_A$, $W_{NA}$ at ${\cal O}(1/c^2)$.
Among these operators,
$V_{\rm QCD}(r)$ (after cancelling the order $\Lambda_{\rm QCD}$
renormalon) and $W_{NA}$ contain order $\Lambda_{\rm QCD}^3 r^2$
renormalons.
Therefore, the order $\Lambda_{\rm QCD}^3 r^2$
renormalons do not cancel completely
in the perturbative computation of the $S$--$P$ splittings.
Namely, the theoretical uncertainties of the $S$--$P$ splittings are
expected to be larger than those of
the fine splittings.

\begin{table}
\begin{center}
\begin{tabular}{|c|c|c|c|c|c|c|}
\hline 
\multirow{2}{.3cm}{$\mu$} &
 \multicolumn{6}{|c|}{$2S-1P_1$} \\
& \parbox{1cm}{\centering $E^{(0)}_\psi$} & $U$ & $U\!+\!\delta U$
& \parbox{1.1cm}{\centering $W_A$} & $W_{NA}$ & $W_{NA}\!+\!\delta W_{NA}$\\
\hline
1.0 & 109.4 & 22.1 & 17.6 & $-$10.6 & $-$66.5 &  25.6 \\
2.0 & 110.7 & 15.0 & 15.1 & $-$10.0 & $-$28.9 & $-$19.1 \\
3.0 & 111.1 & 12.7 & 13.7 &  $-$9.8 & $-$20.4 & $-$20.3 \\
4.0 & 111.4 & 11.5 & 12.8 &  $-$9.7 & $-$16.5 & $-$19.4 \\
5.0 & 111.5 & 10.7 & 12.2 &  $-$9.7 & $-$14.2 & $-$18.3 \\
\hline
\end{tabular}
\begin{tabular}{|c|c|c|c|c|c|c|}
\hline 
\multirow{2}{.3cm}{$\mu$} &
  \multicolumn{6}{|c|}{$3S-2P_1$}\\
& \parbox{1cm}{\centering $E^{(0)}_\psi$} & $U$ & $U\!+\!\delta U$
& \parbox{1.1cm}{\centering $W_A$} & $W_{NA}$ & $W_{NA}\!+\!\delta W_{NA}$\\
\hline
1.0 & 60.7 & 12.6 & 9.2 & $-$0.5 & $-$29.8 & 18.8 \\
2.0 & 59.4 &  8.3 & 8.2 & $-$1.5 & $-$13.0 & $-$6.8 \\
3.0 & 59.0 &  7.0 & 7.5 & $-$1.9 &  $-$9.2 & $-$8.2 \\
4.0 & 58.8 &  6.3 & 7.0 & $-$2.1 &  $-$7.5 & $-$8.1 \\
5.0 & 58.7 &  5.9 & 6.7 & $-$2.2 &  $-$6.5 & $-$7.8 \\
\hline
\end{tabular}
\caption{\small
$S$--$P$ splittings. The splittings due to the differences 
in $E^{(0)}_\psi$, $U$, $U\!+\!\delta U$, $W_A$, $W_{NA}$ and 
$W_{NA}\!+\!\delta W_{NA}$ are given for the $2S$--$1P_1$ and $3S$--$2P_1$ 
splittings. All values are in MeV ($\mu$ in GeV).\label{newtab1}}
\end{center}
\end{table}
In Tab.~\ref{newtab1} we show the 
$S$--$P$ splittings of our zeroth--order energy
levels $E^{(0)}_\psi$, which contain the effects of $V_{\rm QCD}(r)$
up to ${\cal O}(1/c^2)$.
The contributions of the operators 
$U$, $W_A$ and $W_{NA}$ to the $S$--$P$ splittings are also
displayed in the same table.
(A convenient formula for evaluating the expectation values of $W_A$
is given in the Appendix.)
We expect that the contributions of these operators
would be smaller than the $E^{(0)}_\psi$--splittings,
since the operators are ${\cal O}(1/c^2)$, whereas $E^{(0)}_\psi$ contains
the ${\cal O}(1/c)$ effects of $V_{\rm QCD}(r)$.
One sees that this expectation is satisfied in most cases,
the only operator giving a contribution comparable 
in magnitude to $E^{(0)}_\psi$  
is $W_{NA}$ for relatively low scales, $\mu \simeq 1$~GeV, 
where this contribution becomes particularly large.
In fact, the scale-dependence of the contribution of $W_{NA}$ is
large, because $W_{NA}$ is proportional to
$\alpha_S^{(4)}(\mu)^2$, whereas $U$ and $W_A$ are proportional only to
$\alpha_S^{(4)}(\mu)$.
We see that the scale--dependence is reduced
by including the effect of the higher--order correction
$\delta W_{NA}$, although $W_{NA}+\delta W_{NA}$ is still
unstable around $\mu = 1$~GeV.
If we choose a relatively large scale $\mu \simeq 3$~GeV,
the contributions of the higher-order corrections,
$\delta U$ and $\delta W_{NA}$, become small
and the scale dependences of the $S$--$P$ splittings are relatively small;
see Tabs.~\ref{newtab1} and \ref{newtab2}.
\begin{table}
\begin{center}
\begin{tabular}{|c|c|c|c|c|c|c|}
\hline 
\multirow{2}{.3cm}{$\mu$} &
 \multicolumn{3}{|c|}{$2S-1P_1$} &  \multicolumn{3}{|c|}{$3S-2P_1$}\\
&$E^{(0)}_\psi$&$+(U, W_A, W_{NA})$&$+(\delta U, \delta W_{NA})$
&$E^{(0)}_\psi$&$+(U, W_A, W_{NA})$&$+(\delta U, \delta W_{NA})$\\
\hline
 1.0 & 109.4 & 54.4 & 141.9 & 60.7 & 42.9 & 88.2\\
 2.0 & 110.7 & 86.8 &  96.8 & 59.4 & 53.2 & 59.3\\
 3.0 & 111.1 & 93.6 &  94.8 & 59.0 & 54.9 & 56.5\\
 4.0 & 111.4 & 96.6 &  95.1 & 58.8 & 55.6 & 55.7\\
 5.0 & 111.5 & 98.3 &  95.7 & 58.7 & 55.9 & 55.4\\
\hline
Exp.\ & \multicolumn{3}{|c|}{130} & \multicolumn{3}{|c|}{100}\\
\hline
\end{tabular}
\caption{\small $S$--$P$ splittings. This table is similar to Tab.~\ref{newtab1},
but here we add up the matrix elements of all the operators contributing
to the $S$--$P$ splitting. For comparison, the first of the three columns for each
level splitting again gives the splitting due to the difference in $E^{(0)}_\psi$.
In the second column additionally $U, W_A$ and $W_{NA}$ have been taken
into account and finally in the third column $\delta U$ and $\delta W_{NA}$
are added. \label{newtab2}}
\end{center}
\end{table}

\begin{figure}
\begin{center}
\psfrag{R}{$r\,[{\rm GeV}^{-1}]$}
\psfrag{WFWNA}
{\hspace{-10mm}$- r^2\left|R_\psi(r)\right|^2 (\delta)W_{NA}$~~~[GeV$^2$]}
\psfrag{WNA}{$W_{NA}$}\psfrag{dWNA}{$\delta W_{NA}$}
\includegraphics[width=12cm]{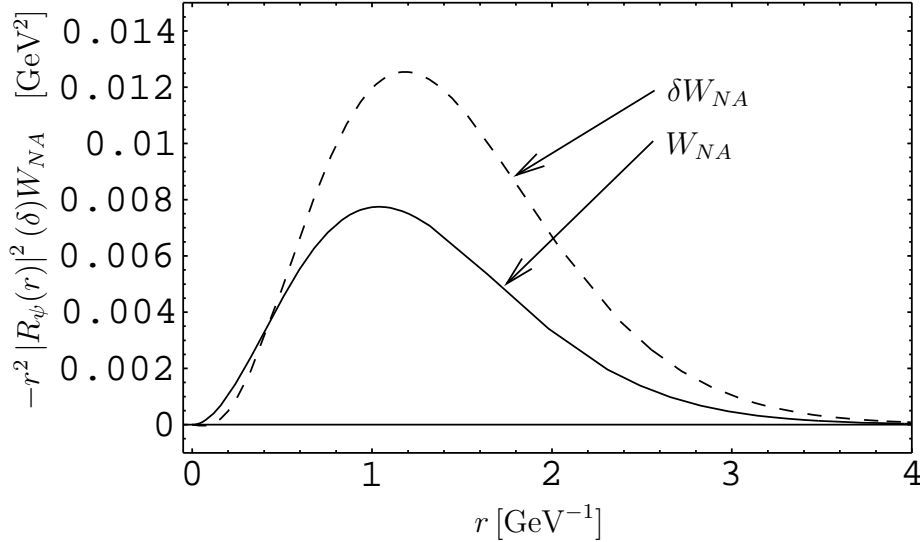}
\end{center}
\vspace*{-.5cm}
\caption{\small
(Absolute value of the) Integrand of the matrix elements of $W_{NA}$ (solid) and
$\delta W_{NA}$ (dashed) for the $1P_j$ states. The graphs peak at rather small
distances, indicating that the natural scales for the expectation values
are larger than those for the boundstate.
\label{newfig1}}
\end{figure}                               
Furthermore,
as shown in Fig.~\ref{newfig1}, the expectation values
$\bra{\psi} W_{NA} \ket{\psi}$ and
$\bra{\psi} W_{NA} + \delta W_{NA} \ket{\psi}$ 
are dominated by short--distance contributions.
Therefore, following 
the same line of argument as in the previous section, we 
expect the natural scales for these contributions to be 
larger than that for the binding energy.

If we compare our predictions for
the $S$--$P$ splittings with the predictions of the
fixed--order perturbative expansion in \cite{bsv2},
there are two competing effects, just like what we found
in the case of the fine splittings: these effects are
the difference of the wave functions and the difference of the scales
$\mu$ in the operator.
Consequently our predictions for the 
$S$--$P$ splittings turn out to be larger for a smaller scale $\mu$,
are typically larger than the fixed--order predictions for
the $2S$--$1P$ splittings, and are of similar magnitude 
to the fixed--order predictions for the
$3S$--$2P$ splittings.
Since, however, the $S$--$P$ splittings are dominated by the
${\cal O}(1/c)$ correction from $V_{\rm QCD}(r)$ [i.e.\ the
contributions from the ${\cal O}(1/c^2)$ operators are only subleading],
the differences from the fixed--order predictions are not as 
pronounced as in the case of the fine splittings.

As long as we calculate the $S$--$P$ splittings directly,
as done up to here,
we see no indication of large theoretical uncertainties.
However, when we examine the individual energy levels,
some indications of fairly large uncertainties show up.
Let us now investigate this feature.

\begin{table} \begin{center}
\begin{tabular}{|c|c|c|c|c|c|c|c|c|c|}
\hline
& $1S$ & $1P_0$ & $1P_1$ & $1P_2$ & $2S$ & $2P_0$ & $2P_1$ & $2P_2$ & $3S$ \\
\hline 
$U$              &  13.2 & $-$25.7 &  $-$6.4 &   9.0 &   6.3 & $-$15.4 &  $-$3.8 &   5.4 &   3.2 \\
$U\!+\!\delta U$ &  13.2 & $-$30.7 &  $-$7.4 &  10.6 &   6.3 & $-$17.9 &  $-$4.3 &   6.2 &   3.2 \\
$W_A$            & $-$22.1 & $-$18.8 & $-$18.8 & $-$18.8 & $-$28.6 & $-$18.9 & $-$18.9 & $-$18.9 & $-$20.8 \\
$W_{NA}$         & $-$79.1 & $-$11.9 & $-$11.9 & $-$11.9 & $-$32.3 &  $-$6.4 &  $-$6.4 &  $-$6.4 & $-$15.6 \\
$W_{NA}\!+\!\delta W_{NA}$ 
                & $-$130.2 & $-$31.4 & $-$31.4 & $-$31.4 & $-$51.7 & $-$16.8 & $-$16.8 & $-$16.8 & $-$25.0 \\
\hline
\end{tabular} \end{center}
\caption{\small
Expectation values of the operators 
$U$, $U\!+\!\delta U$, $W_A$, $W_{NA}$ and $W_{NA}\!+\!\delta W_{NA}$ at
the scale $\mu=3$~GeV. \label{newtab3}}
\end{table}
Tab.~\ref{newtab3} shows
the expectation values of the operators 
$U$, $U \!+\! \delta U$, $W_A$, $W_{NA}$, $W_{NA} \!+\! \delta W_{NA}$
for all the states and with $\mu=3$~GeV.
We see that the expectation values of $W_{NA}$ and 
$W_{NA} \!+\! \delta W_{NA}$ for the $1S$ state
are much larger than what we would expect for 
${\cal O}(1/c^2)$ corrections.
Moreover, for all the states, 
the scale--dependences of the expectation values of
$W_{NA} \!+\! \delta W_{NA}$ are large, and are comparable to
those of  $W_{NA}$; see Tab.~\ref{newtab4}.
\begin{table} \begin{center}
\begin{tabular}{|c|c|c|c|c|c|c|c|c|c|}
\hline 
\multirow{2}{.3cm}{$\mu$} &
 \multicolumn{4}{|c|}{$2S$} & \multicolumn{5}{|c|}{$1P_1$}\\
& $U$ & $W_A$ & $W_{NA}$ & $W_{NA}\!+\!\delta W_{NA}$
& $U$ & $U+\delta U$ & $W_A$ & $W_{NA}$ & $W_{NA}\!+\!\delta W_{NA}$\\
\hline
1.0 & 11.0 & $-$36.7 & $-$104.8 & $-$73.0 & $-$11.1 & $-$6.5 & $-$26.1 & $-$38.3 & $-$98.6 \\ 
2.0 &  7.4 & $-$30.6 &  $-$45.6 & $-$63.0 &  $-$7.6 & $-$7.7 & $-$20.5 & $-$16.8 & $-$44.0 \\
3.0 &  6.3 & $-$28.6 &  $-$32.3 & $-$51.7 &  $-$6.4 & $-$7.4 & $-$18.8 & $-$11.9 & $-$31.4 \\
4.0 &  5.7 & $-$27.5 &  $-$26.2 & $-$45.0 &  $-$5.8 & $-$7.2 & $-$17.8 &  $-$9.6 & $-$25.7 \\
5.0 &  5.3 & $-$26.9 &  $-$22.6 & $-$40.6 &  $-$5.4 & $-$7.0 & $-$17.2 &  $-$8.3 & $-$22.3 \\
\hline
\end{tabular} \end{center}
\caption{\small
Expectation values of the operators 
$U$, $U\!+\!\delta U$, $W_A$, $W_{NA}$ and $W_{NA}\!+\!\delta W_{NA}$ for
the states $2S$ and $1P_1$ for various scales $\mu$. For the $2S$ state
we leave out $U\!+\!\delta U$ because $\delta U=0$ for $l=0$. \label{newtab4}}
\end{table}
(A large part of these scale--dependences cancel in the 
$S$--$P$ splittings.) The reason for the large scale--dependences
is that the non--logarithmic term of $\delta W_{NA}$ is very large.
Since at present we do not know the full form of the Hamiltonian
up to ${\cal O}(1/c^3)$, we cannot draw a definitive conclusion 
whether this problem of large scale--dependence can be remedied.

The agreement of the $S$--$P$ splittings between the
theoretical predictions and the experimental data has
improved somewhat from the fixed--order results \cite{bsv2}.
It is, however, not as good as one would naively expect.
Namely the differences of the splittings
between our predictions (for the scale choice $\mu \simeq 2$--5~GeV)
and the experimental data are larger than the contributions of
the ${\cal O}(1/c^2)$ operators.
As stated in the introduction, it is important to clarify whether
the level of disagreement
is still consistent within perturbative uncertainties.
From the examination of the contributions to the
individual energy levels, we conjecture that there would still be
large theoretical uncertainties to the $S$--$P$ splittings, particularly
from the contributions of $W_{NA}$,
since at present we have no systematic argument on how much of the
${\cal O}(\Lambda_{\rm QCD}^3)$ renormalons cancel in the
$S$--$P$ splittings.
That is, we consider the cancellation of the large and 
scale--dependent contributions from $W_{NA}$ in the
$S$--$P$ splittings to be accidental, unless we find
a systematic argument in support of it.
In this regard, we consider the predictions 
for the $S$--$P$ splittings much less
reliable than those for the fine splittings.

There are also other indications that the operator $W_{NA}$
becomes a source of instability of
theoretical predictions.
For instance, ref.~\cite{nos} addresses $W_{NA}$ to be the source of
the large uncertainties of the cross section for  $e^+e^- \to t\bar{t}$
close to threshold.
Ref.~\cite{n3lo} shows that IR logarithms, which are related to
higher--order corrections to $W_{NA}$ and $V_{\rm QCD}$, 
generate very large corrections to the bottomonium spectrum
at ${\cal O}(\alpha_S^5 m_b)$
and consequently cause a large uncertainty to it.
Furthermore, 
refs.~\cite{ams,man-stew,hmst} resum the IR logs; 
it was found that $W_{NA}$ is reduced due to this resummation
in the corresponding Wilson coefficients.
Ref.~\cite{hmst} claims that the resummations
lead to a stable theoretical prediction for the top threshold cross section.
Ref.~\cite{p02}, however, reported a result which disagrees with
that of \cite{ams,man-stew}.

In this paper we discard the IR logs altogether, which appear first
at ${\cal O}(\alpha_S^5 m_b \log \alpha_S)$ in the spectrum.
This is not 
because we consider them unimportant, but because we believe that we are 
not yet in a position to treat this problem properly.  
The present treatment of the IR logs in \cite{kpss1}, \cite{ams}--\cite{p02}
seems to comprise following unsatisfactory aspects
in the calculations of the heavy quarkonium spectrum and
the top threshold cross section (besides the disagreement).
Since $\log \alpha_S$ is not particularly large, it appears that
the non--logarithmic terms cannot be neglected 
in comparison to the IR logs.
These non--logarithmic terms are not yet fully known at 
${\cal O}(\alpha_S^5 m_b )$ and beyond, so that we are unable to
incorporate them unambiguously.
Furthermore, if a resummation of the IR logs 
stabilizes the theoretical predictions substantially, we would like
to understand the physical meaning behind it, i.e.\
why the resummation is important.
A hint to these questions was suggested in \cite{offshell}:
it shows that 
${\cal O}(\Lambda_{\rm QCD}^3)$ renormalons will be suppressed if we
incorporate the offshellness of the quarks, which acts as an
IR cutoff in the temporal dimension.
This applies to the 
${\cal O}(\Lambda_{\rm QCD}^3)$ renormalons contained in
$W_{NA}$ and $V_{\rm QCD}$ (and their higher--order corrections),
and they are closely related to the IR logs and their resummation.
We suspect that 
further investigations of all these problems may be a way to
clarify and solve the problem of the operator $W_{NA}$
we face here.

\section{The Spectrum}

\begin{figure}
\begin{center}
\psfrag{BSV}{BSV}\psfrag{1181}{(1181)}\psfrag{1161}{(1161)}
\psfrag{EQuigg}{Eichten--}\psfrag{EQuigg2}{Quigg}
\psfrag{Exp}{Exp}
\psfrag{1.}{$\mu=1$}\psfrag{2.}{$\mu=2$}\psfrag{3.}{$\mu=3$}
\psfrag{4.}{$\mu=4$}\psfrag{5.}{$\mu=5$}
\includegraphics[width=15cm]{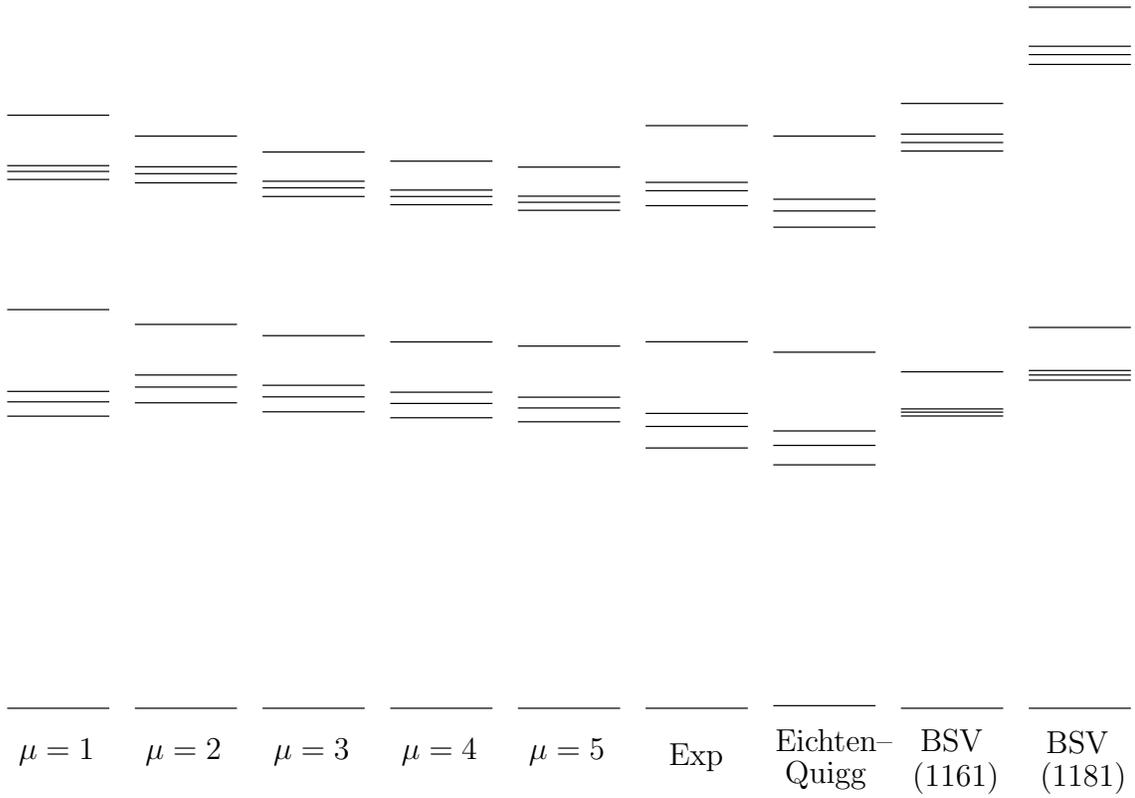}
\end{center}
\vspace*{-.5cm}
\caption{\small Comparison of the energy levels obtained with different
formalisms. The columns labelled $\mu=1$ through $\mu=5$ show our results, 
where now $\alpha_S^{(5)}(M_Z) = 0.1181$ and $\overline{m}_b=4.234$~GeV
have been used to make the $1S$ state coincide with experiment
for $\mu=3$~GeV. The columns for $\mu=1$, 2, 4 and 5~GeV have been
shifted to achieve this coincidence. ``Exp'' shows the experimental values
and ``Eichten--Quigg'' those obtained in \cite{eq}.
Finally, ``BSV'' corresponds to the formalism of \cite{bsv2},
the two columns represent choices of 
$\alpha_s^{(5)}(M_Z)=0.1161$ and $0.1181$, 
respectively. \label{spectrum}}
\end{figure}

In this section we compare the
whole bottomonium spectrum as determined experimentally
with various theoretical predictions. We list the energy levels
numerically in Tab.~\ref{energylevels} and show the spectrum
in Fig.~\ref{spectrum}. The levels were calculated according to 
the framework explained in Sec.~\ref{s2}, using the input parameter 
$\alpha_S^{(5)}(M_Z) = 0.1181$ and including the effects of 
$\delta U$ and $\delta W_{NA}$. We employed the scale choices 
$\mu = 1$, 2, 3, 4 and 5~GeV.
\begin{table} \begin{center}
\begin{tabular}{|c|c|c|c|c|c|c|c|c|c|}
\hline
$\mu$ & $1S$ & $1P_0$ & $1P_1$ & $1P_2$ & $2S$ & $2P_0$ & $2P_1$ & $2P_2$ & $3S$ \\
\hline 
1.0 & 9.369 & 9.818 & 9.840 & 9.856 &  9.982 & 10.182 & 10.194 & 10.203 & 10.281 \\
2.0 & 9.424 & 9.894 & 9.918 & 9.936 & 10.014 & 10.231 & 10.245 & 10.256 & 10.303 \\
3.0 & 9.460 & 9.915 & 9.938 & 9.956 & 10.032 & 10.246 & 10.260 & 10.270 & 10.315 \\
4.0 & 9.481 & 9.927 & 9.949 & 9.966 & 10.043 & 10.254 & 10.267 & 10.277 & 10.321 \\
5.0 & 9.494 & 9.934 & 9.955 & 9.972 & 10.050 & 10.259 & 10.271 & 10.281 & 10.325 \\
\hline
Exp.& 9.460 & 9.860 & 9.893 & 9.913 & 10.023 & 10.232 & 10.255 & 10.268 & 10.355 \\
\hline
\end{tabular} \end{center}
\caption{\small
Energy levels for all states including $E^{(0)}_\psi$, 
$U\!+\!\delta U$, $W_A$, and $W_{NA}\!+\!\delta W_{NA}$. The parameters
used are $\alpha_S^{(5)}(M_Z) = 0.1181$ and $\overline{m}_b=4.234$~GeV,
this choice of $\overline{m}_b$ is explained in the text.
\label{energylevels}}
\end{table}

In this section we use $\overline{m}_b=4.234$~GeV instead of
$\overline{m}_b=4.190$~GeV to make the the prediction for the $1S$ state
coincide with the experimental value for our favoured scale
of $\mu = 3$~GeV.\footnote{
The change of $\overline{m}_b$ by this adjustment does not alter the
qualitative features of the predictions for the
fine splittings and the $S$--$P$ splittings discussed in
previous sections.
}.
We consider this different choice of $\overline{m}_b$ a more natural
way to achieve coincidence with the experimental value than a
simple shift of the whole spectrum; the numerical difference between 
these two prescriptions is actually very minor.

In Fig.~\ref{spectrum} we compare our results to the spectrum obtained 
from the experimental data (``Exp''), the predictions of the
fixed--order perturbative expansions (``BSV''),\footnote{
We follow the scheme and the scale--fixing condition A of Sec.~4.2 of
\cite{bsv2}, except that we
use the numerical solution to the renormalization--group
equation for the strong coupling constant.
The results are obtained for the input parameters
$\alpha_S^{(5)}(M_Z) = 0.1181$ and 0.1161; the prediction with the
latter input agrees better with the experimental data.
}
and the result of \cite{eq} as a typical prediction of 
recent phenomenological models (``Eichten-Quigg'').

One can verify the conclusions of our analysis in the previous
two sections:
The fine splittings and the $S$--$P$ splittings are larger than those of
the fixed-order results for the $n=2$ states, whereas they are of
similar magnitude to the fixed-order results for the $n=3$ states.
The scale-dependence of our predictions originates mostly from the
scale--dependence of the operator $W_{NA} + \delta W_{NA}$; c.f.\ the 
discussion in the last section.
Only the gross level spacings between adjacent $n$'s are affected
visibly by changes of $\mu$ between 2--5~GeV, whereas
also the $S$--$P$ splittings vary visibly
for a smaller $\mu$ between 1--2~GeV.
The size of this variation is very large considering that the
scale dependence is formally an ${\cal O}(1/c^3)=O(\alpha_S^5m_b)$
effect.
The level spacings between consecutive $n$'s 
as well as the $S$--$P$ splittings increase for smaller $\mu$.
We regard the large scale-dependence generated by $W_{NA}$
and its higher-order corrections
as the largest theoretical uncertainty of our prediction.  

Let us note the effects of the operator $W_{NA}$ or
$W_{NA} + \delta W_{NA}$ in particular:
the level spacings between consecutive $n$'s are increased,
while the $S$--$P$ splittings are reduced.
The reason can be understood as follows.
The operator $W_{NA} (+ \delta W_{NA})$ generates an attractive potential
proportional to $1/r^2$ (with a logarithmic correction)
which is particularly strong at short distances.
Hence, those states which have larger wave functions close to
the origin acquire larger binding energies.
Since the states with lower $n$ have larger amplitude close to
the origin and therefore
acquire larger binding energies,
the level spacings between the adjacent $n$'s become wider.
Since the $S$ states acquire larger binding energies than the $P$ states,
the $S$--$P$ splittings becomes narrower.

Generally, our prediction of the spectrum has a better agreement with
the experimental data than the fixed--order results.
The agreement seems to be better for a larger scale choice, which
appears reasonable, since the natural scale of the
operator $W_{NA} + \delta W_{NA}$ would be large.
(The scale dependences due to other effects are much smaller.)
The agreement seems to be optimal for a scale choice
$\mu \simeq 3$--4~GeV.
We also examined our predictions for different values of the
input $\alpha_S^{(5)}(M_Z)$ within the present world--average values
$0.1181 \pm 0.0020$ \cite{pdg}.
We find that generally all the level spacings and splittings become
larger for larger $\alpha_S^{(5)}(M_Z)$, since the binding energy
increases.
The widening of the level spacings, however, can be compensated
largely by choosing a larger value for $\mu$.
The spectrum of the
phenomenological model still has a better agreement with the experimental
data, since it includes more parameters which can be adjusted.

In general one should carefully take into account
theoretical uncertainties
when comparing the whole spectrum with the experimental data.
Based on the renormalon argument, each energy level has (at least)
an uncertainty of order $\Lambda_{\rm QCD}^3 r^2$, where
$r$ should be taken as a typical size of the quarkonium state.
Theoretical uncertainties contained in $E_{\rm imp}(r)$ in the
region $r < r_{\rm IR}$ can be represented typically by these
renormalon estimates \cite{paper1,RS1}.
Ref.~\cite{bsv2} estimated the uncertainties to be
$\pm(5$--$30)$ MeV for the $1S$ state, $\pm(20$--$130)$ MeV
for the $n=2$ states, and $\pm(40$--$220)$ MeV for the $n=3$ states.
On the other hand, 
the level spacings (splittings) have smaller theoretical uncertainties
since these theoretical uncertainties cancel, at least partly,
as we discussed in the previous two sections.

\begin{figure}
\begin{center}
\psfrag{-300}{$m_b\!-\!300$~MeV}\psfrag{+300}{$m_b\!+\!300$~MeV}\psfrag{QCD}{QCD}
\psfrag{half}{${1\over2} \, C_1$}
\psfrag{double}{$2 \, C_1$}
\includegraphics[width=14cm]{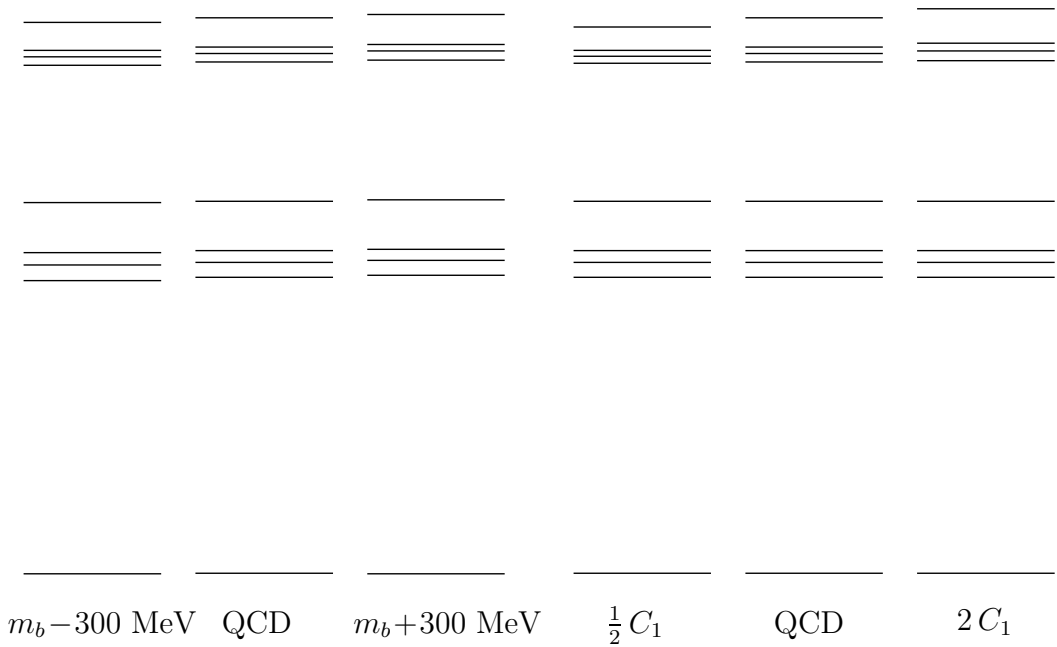}
\end{center}
\caption{\small
Analysis of various uncertainties: In the left part of the diagram 
we show the effect of a shift of $\pm 300$~MeV of the pole mass $m_b$ while
keeping the $\overline{\rm MS}$ mass $\overline{m}_b$ that enters the potential
$E_{\rm imp}(r)$
constant. To make the $1S$ states concide, we have shifted the spectrum
for $m_b\!-\!300$~MeV ($m_b\!+\!300$~MeV) down (up) by about 20~MeV. The
right part shows the effect of changing the slope of the IR part of
$E_{\rm imp}(r)$ by a factor of $1/2$ (2). 
(The definition of $C_1$ is given in Sec.~2.2.) 
The effect of this change on the $1S$
state is negligible, so that the spectra did not have to be shifted.
\label{ErrorAnalysis}}
\end{figure}                               
Let us estimate the errors of our prediction from other sources.
In the left part of Fig.~\ref{ErrorAnalysis} we show the effects 
of a variation of the $b$--quark pole mass by $\pm 300$~MeV from
the value listed in Tab.~\ref{polemass}. (For the error estimates
we set the scale to $\mu=3$~GeV.)
The states with principal quantum numbers $n=1$ and 2 are shifted
up (down) by about $20$~MeV when the pole mass is shifted down
(up) by 300~MeV; for the states with principal quantum number $n=3$
this variation is about $\pm 15$~MeV.
Consequently if we compensate the overall shift such that the
$1S$ level agrees with the experimental value, only the $n=3$
levels vary by about $\pm 5$~MeV.

In the right part of Fig.~\ref{ErrorAnalysis} 
we also show the effect of a variation of the long--distance part 
of the potential $E_{\rm imp}(r)$ in the zeroth--order
Schr\"odinger equation.
We vary the tangent of the linear potential at $r > r_{\rm IR}$
by factors of 2 and 1/2.
(The first derivative of the potential then becomes discontinuous
at $r = r_{\rm IR}$.)
The variations of the energy levels are of order $\pm(2$--$5)$~MeV 
for the $2P$ states and $\pm 15$~MeV for the $3S$ state,
while they are smaller than $0.1$~MeV (and therefore not visible in 
Fig.~\ref{ErrorAnalysis}) for the lower states. This can be easily
understood because only the states with principal quantum number
$n=3$ have a wave function that extends to large enough distances
to probe the potential in this region.

\begin{figure}
\begin{center}
\psfrag{BSV}{BSV}\psfrag{1181}{(1181)}\psfrag{1161}{(1161)}
\psfrag{EQuigg}{Eichten--}\psfrag{EQuigg2}{Quigg}
\psfrag{Exp}{Exp}
\psfrag{1.}{$\mu=1$}\psfrag{2.}{$\mu=2$}\psfrag{3.}{$\mu=3$}
\psfrag{4.}{$\mu=4$}\psfrag{5.}{$\mu=5$}
\includegraphics[width=15cm]{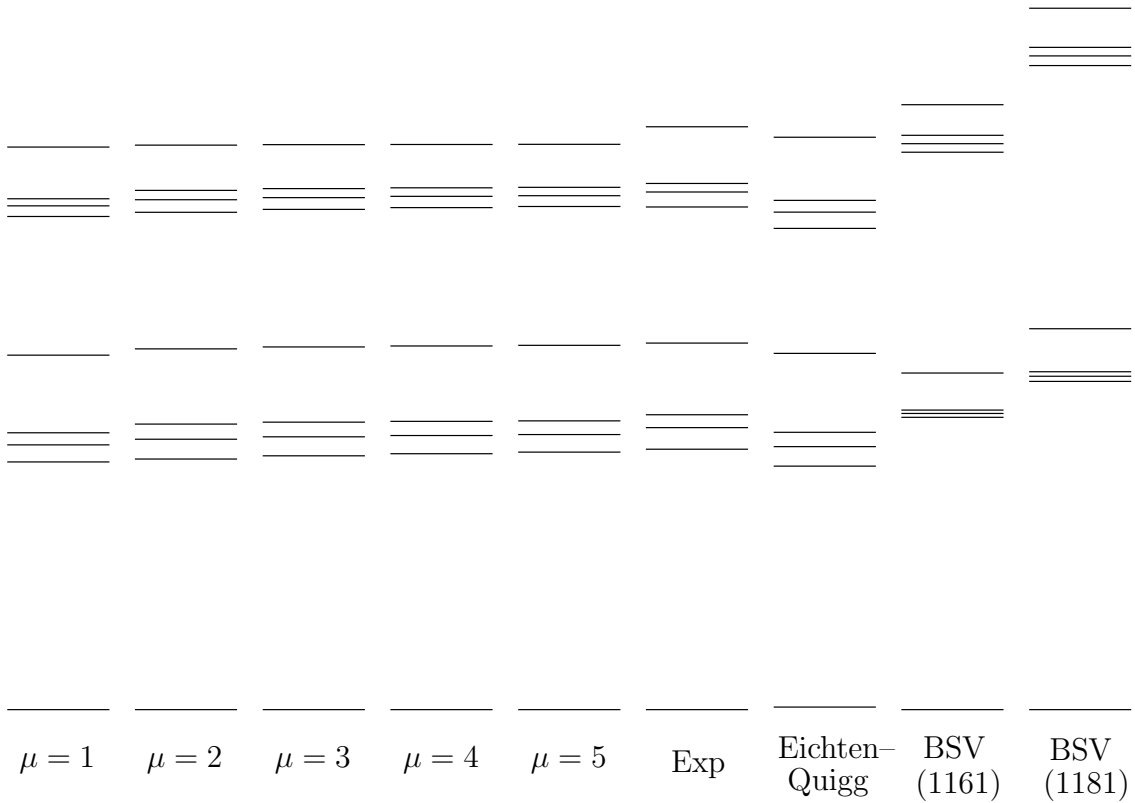}
\end{center}
\vspace*{-.5cm}
\caption{\small
For comparison we show the bottomonium
spectrum with the $W_{NA} + \delta W_{NA}$ term artificially
set to 0 and $\alpha_S^{(5)}(M_Z) = 0.1201$
and $\overline{m}_b=4.151$~GeV, otherwise the 
conventions are as in  Fig.~\ref{spectrum}. The scale dependence 
is decreased and the agreement with the experimental
spectrum is strongly improved with respect to Fig.~\ref{spectrum},
indicating a phenomenological preference for a suppression
mechanism for $W_{NA}$.\label{noWNA}}
\end{figure}

For the sake of comparison, we show in Fig.~\ref{noWNA} the 
predictions for the bottomonium spectrum when we set the
$W_{NA} \!+\! \delta W_{NA}$ to 0 artificially (in this figure
we use $\alpha_S^{(5)}(M_Z) = 0.1201$ and $\overline{m}_b=4.151$~GeV).
We see that the scale--dependence reduces 
(as expected) and there is a much better agreement with the experimental data.
Phenomenologically this may be taken as an indication that
the contributions of $W_{NA}$ should be suppressed by some mechanism.

\section{Conclusions}

We have examined the bottomonium spectrum within a specific
framework based on perturbative QCD.
The computation of the individual energy levels includes all the effects up to
${\cal O}(\alpha_S^4 m_b)$ and that of the fine splittings contains
all the effects up to ${\cal O}(\alpha_S^5 m_b)$.
We have also included some important higher--order 
corrections to the quarkonium wave functions and the energy levels
through our use of $E_{\rm imp}(r)$,
which is the characterizing feature of our analysis.
We have eliminated the scale--dependence from $E_{\rm imp}(r)$ by
fixing a specific scheme and we have discussed scale--depedences 
originating from the other terms of the Hamiltonian.
The agreement of the
fine splittings among the $1P_j$ states between the theoretical
prediction and the experimental data improved drastically as compared
to the fixed--order prediction of \cite{bsv1,bsv2}.
We find that the centering of the wave functions 
towards the origin as compared to
the Coulomb wave functions, due to the strong
attractive force in the intermediate--distance region, strongly enhances the
fine splittings.
We also find that, in accord with a na\"\i ve expectation, the natural scales of 
the fine splittings 
are larger than those of the boundstates themselves;
the latter were used in the analyses of \cite{bsv1,bsv2}.
The predictions for the fine splittings are stable against the
variation of the scale $\mu$ and are in reasonable agreement with
the experimental data both for the $1P_j$ and $2P_j$ states.
We also examined the $S$--$P$ level splittings.
The agreement with the experimental data has improved
as compared to the fixed--order results, but the predictions
are still somewhat smaller than the experimental values.
The predictions for these 
splittings are stable against the variation of $\mu$
between 2--5~GeV but become unstable for lower scales
between 1--2~GeV.
Natural scales of the $S$--$P$ splittings 
are also found to be
larger than those of the boundstates.
On the other hand,
the predictions of the level spacings between the
adjacent $n$'s depend rather strongly on the scale $\mu$.
This stems from the large scale dependence of the operator
$W_{NA}$ and must be regarded as a major source of 
uncertainties in our predictions.
We are motivated to choose
a relatively large value for the scale $\mu$ in view of
the dominance of short-distance contributions to
$\bra{\psi} W_{NA} \ket{\psi}$. 
(Other effects are much less scale dependent.)
If we choose $\mu \simeq 3$--4~GeV, and for
$\alpha_S^{(5)}(M_Z) = 0.1181$ and $\overline{m}_b=4.234$~GeV, the
agreement between our prediction and the experimental data for
the whole bottomonium spectrum
is fairly good, and is considerably better than 
the agreement between the fixed--order prediction and the experimental
data.
There seem to be some indications, however, 
that the contribution of the operator
$W_{NA}$ reduces the stablity of the theoretical prediction and
at the same time
worsens the agreement between the prediction and the experimental data.
At the present state we consider
our predictions to be consistent with
the experimental data within theoretical uncertainties. 

\section*{Appendix}

We derive a formula which is
convenient for evaluating the expectation value of
$W_A$ [Eq.~(\ref{WA})] with respect to the eigenstate $\ket{\psi}$
of $H_0^{\rm (imp)}$ defined in Eq.~(\ref{schroedinger-eq}).
We substitute the following operator identities to $W_A$:
\bea
&&
\frac{1}{r^3} \, r^i r^j p^j p^i =
\vec{p}\,^2 \, \frac{1}{r}  - \frac{\vec{L}^2}{r^3}
- 4 \pi \, \delta^{(3)}(\vec{r}) ,
\\
&& 
\vec{p}\,^2 = m_b \, \left[ H_0^{\rm (imp)} - E_{\rm imp}(r) \right] .
\eea
Then one finds
\bea
W_A & =  &- \frac{1}{4 m_b} \left[ H_0^{\rm (imp)} - E_{\rm imp}(r) \right]^2
+
\frac{3\pi C_F \alpha_S}{m_b^2} \, \delta^{(3)}(\vec{r})
- \frac{C_F \alpha_S}{2m_b} \,
\left\{ \frac{1}{r}, \, H_0^{\rm (imp)} - E_{\rm imp}(r) \right\}
\nonumber  \\ &&
+ \frac{C_F \alpha_S}{2 m_b^2} \, \frac{\vec{L}^2}{r^3} .
\eea
Hence, the expectation value can be written as
\bea
\bra{\psi} W_A \ket{\psi} & = &
- \frac{1}{4 m_b} 
\left\langle \left[ E_\psi^{(0)} - E_{\rm imp}(r) \right]^2 \right\rangle
- \frac{C_F \alpha_S}{m_b} \,
\left\langle \frac{ E_\psi^{(0)} - E_{\rm imp}(r) }{r} \right\rangle
\nonumber  \\ &&
+ \frac{C_F \alpha_S}{2 m_b^2} \, l(l+1) \,
\left\langle \frac{1}{r^3}\right\rangle
+
\frac{3\pi C_F \alpha_S}{m_b^2} \, |\psi(\vec{0})|^2 .
\eea
All quantities on the right-hand-side can be evaluated from
the radial wave function and the energy eigenvalue, which 
are obtained by solving the 
Schr\"odinger equation numerically.

\section*{Acknowledgements}
Y.S. is grateful to Y.~Kiyo for a valuable suggestion.
S.R. was supported by the Japan Society for the Promotion of Science
(JSPS).
\vspace{3mm}\\

\noindent
{\bf \large Note Added in Proof:}\vspace{2mm}\\
After this paper was submitted, we received \cite{new} which
confirmed in vNRQCD framework the result of pNRQCD \cite{p02},
and the disagreement we mentioned at the end of Sec.~5 has been
resolved.

\end{document}